\def\year{2022}\relax
\documentclass[letterpaper]{article} 
\usepackage{aaai22}  
\usepackage{times}  
\usepackage{helvet}  
\usepackage{courier}  
\usepackage[hyphens]{url}  
\usepackage{graphicx} 
\usepackage{natbib}  
\usepackage{caption} 
\DeclareCaptionStyle{ruled}{labelfont=normalfont,labelsep=colon,strut=off} 
\usepackage{algorithm}
\usepackage[noend]{algorithmic}
\usepackage{booktabs}
\usepackage{amsthm}
\usepackage{thmtools, thm-restate}

\usepackage{newfloat}
\usepackage{listings}
\floatstyle{ruled}
\newfloat{listing}{tb}{lst}{}
\floatname{listing}{Listing}

\usepackage[utf8]{inputenc} 
\usepackage[T1]{fontenc}    
\usepackage{booktabs}       
\usepackage{amsfonts}       
\usepackage{nicefrac}       
\usepackage{float}
\usepackage{multicol}
\usepackage{multirow}
\usepackage{siunitx}
\usepackage{tikz}
\usepackage{pgf}
\usepackage{mleftright}
\usepackage{enumitem}

\usetikzlibrary{calc,shapes,arrows,positioning,arrows.meta}

\title{Fast Payoff Matrix Sparsification Techniques for Structured Extensive-Form Games}

\author{
  Gabriele Farina\textsuperscript{\rm 1},
  Tuomas Sandholm\textsuperscript{\rm 1,2,3,4}
}
\affiliations{
  \textsuperscript{\rm 1}  Department of Computer Science,
  Carnegie Mellon University,
  Pittsburgh, PA 15213 \\
  \textsuperscript{\rm 2} Strategy Robot, Inc.\\
  \textsuperscript{\rm 3} Optimized Markets, Inc.\\
  \textsuperscript{\rm 4} Strategic Machine, Inc.\\
  \{gfarina,sandholm\}@cs.cmu.edu
}

\usepackage{bm}
\usepackage{natbib}
\usepackage{amsmath}
\usepackage{amsthm}
\usepackage{amssymb}
\usepackage{mathtools}
\usepackage{xintexpr}

\usepackage[capitalise,noabbrev]{cleveref}
\usepackage{xcolor}

\renewcommand{\vec}[1]{\bm{#1}}
\newcommand{\kro}{\otimes}

\newcommand{\bbR}{\mathbb{R}}

\newcommand{\n}[1]{%
  \xintifLt{#1}{1}{%
      \num{\xinttheiexpr{#1 * 1000}\relax}ms%
  }{%
    \xintifLt{#1}{60}{%
      \num[round-mode=places,round-precision=2]{#1}s%
    }{
      \xintAssign\xintiiDivision{\xintNum{#1}}{3600}\to\hours\minutes%
      \xintAssign\xintiiDivision{\minutes}{60}\to\minutes\seconds%
      \xintiiifGt{\hours}{0}{
         \num{\hours}h \num[minimum-integer-digits=2]{\minutes}m%
      }{
         \num{\minutes}m \num[minimum-integer-digits=2]{\seconds}s%
      }%
    }%
  }%
}

\newtheorem{proposition}{Proposition}
\newtheorem{property}{Property}
\crefname{property}{Property}{Properties}

\crefname{example}{Example}{Examples}
\newcommand{\defeq}{:=}
\newcommand{\seq}[1]{\textsf{#1}}
\newcommand{\hideseq}[1]{}

\tikzset {
    pl1/.style={%
        semithick,fill=black,draw=black,circle,inner sep=.55mm,
        append after command={
            \pgfextra{
            \begin{pgfinterruptpath}
                \draw[gray] (\tikzlastnode.center) circle (.2);
            \end{pgfinterruptpath}
        }}
    },
    pl2/.style={%
        semithick,fill=white,draw=black,circle,inner sep=.58mm,
        append after command={
            \pgfextra{
            \begin{pgfinterruptpath}
                \draw[gray] (\tikzlastnode.center) circle (.2);
            \end{pgfinterruptpath}
        }}
    },
    showdown/.style={%
        draw=black,inner sep=.9mm,
        append after command={
            \pgfextra{
            \begin{pgfinterruptpath}
                \draw[fill=black] ($(\tikzlastnode.center) - (.25mm,.25mm)$) rectangle +(.5mm,.5mm);
            \end{pgfinterruptpath}
        }}
    },
    fold/.style={%
        draw=black,inner sep=.9mm
    }
}
\newcommand*\circled[1]{%
    \tikz[baseline=(C.base)]\node[draw,circle,inner sep=0.5pt](C) {\normalfont #1};\!
}
\let\xsubsection\subsection
\newcommand*{\daggerfootnote}[1]{%
    \renewcommand*{\thefootnote}{$\dagger$}%
    \footnotetext{#1}%
    \renewcommand*{\thefootnote}{\arabic{footnote}}%
}

\begin{document}
\maketitle

\begin{abstract}
    The practical scalability of many optimization algorithms for large extensive-form games is often limited by the games' huge payoff matrices. To ameliorate the issue, Zhang and Sandholm (2020) recently proposed a sparsification technique that factorizes the payoff matrix $\vec{A}$ into a sparser object $\vec{A} = \hat{\vec{A}} + \vec{U}\vec{V}^\top$, where the total combined number of nonzeros of $\hat{\vec{A}}$, $\vec{U}$, and $\vec{V}$ is significantly smaller. Such a factorization can  be used in place of the original payoff matrix in many optimization algorithm, such as interior-point and second-order methods, thus increasing the size of games that can be handled. Their technique significantly sparsifies poker (end)games, standard benchmarks used in computational game theory, AI, and more broadly.
    We show that the existence of extremely sparse factorizations in poker games can be tied to their particular \emph{Kronecker-product} structure. We clarify how such structure arises and introduce the connection between that structure and sparsification. By leveraging such structure, we give two ways of computing strong sparsifications of poker games (as well as any other game with a similar structure) that are i) orders of magnitude faster to compute, ii) more numerically stable, and iii) produce a dramatically smaller number of nonzeros than the prior technique. Our techniques enable---for the first time---effective computation of high-precision Nash equilibria and strategies subject to constraints on the amount of allowed randomization. Furthermore, they significantly speed up parallel first-order game-solving algorithms; we show state-of-the-art speed on a GPU.

\end{abstract}


\section{Introduction}

Certain important quantities of interest in computational game theory can be expressed as the solution to a linear program (LP) and therefore---in principle---solved for by any algorithm for linear optimization. The practice is more nuanced. The size of the LP is usually dominated by the payoff matrix of the game, that is, the matrix of payoffs for each of the possible terminal states of the game. Correspondingly, in large extensive-form games, most out-of-the-box algorithms for linear programming---such as interior point methods and the simplex method---are unviable. This has historically led the research community in computational game theory to develop specialized (as opposed to applicable to any linear program) algorithms---usually first-order methods---that avoid the need for representing the payoff matrix explicitly. Among these, the most successful examples include the CFR algorithm~\citep{Zinkevich07:Regret} and its modern variants~\citep{Tammelin14:Solving,Moravvcik17:DeepStack,Brown17:Superhuman,Brown17:Reduced,Brown19:Solving,Brown19:Superhuman,Davis19:Solving,Farina21:Faster,Morrill21:Efficient}, and methods based on accelerated first-order methods such as EGT~\citep{Nesterov05:Excessive,Hoda10:Smoothing,Kroer18:Solving,Farina21:Better} and Mirror Prox~\citep{Nemirovski04:Prox,Kroer19:First,Farina21:Better}, which are able to scale to large two-player extensive-form games and compute approximate Nash equilibria for moderate approximation gaps. However, there exist certain applications where currently only LP and linear integer programming technology provide suitable guarantees. For example, the only scalable method for computing sequentially-rational equilibria depends on the ability to find high-precision Nash equilibria, a task that can currently only be achieved using LP technology. Another example is computation of strategies subject to constraints such as  support size, sparsity, or amount of randomization%
, an optimization problem that can easily be expressed via integer linear programming.

In a recent paper, \citet{Zhang20:Sparsified} propose a technique to factorize the payoff matrix of any two-player extensive-form game into a low-rank decomposition (called a \emph{sparsification}) such that the number of nonzeros required in the decomposition is significantly smaller than the number of nonzeros in the original payoff matrix. They show that such a factorization can then be used in place of the original payoff matrix in certain LPs, thereby increasing the game size that LP technology is able to handle. While their sparsification technique is able to typically reduce the number of nonzeros by a factor of 2--3, a notable empirical finding in their evaluation is the dramatic reduction in the number of nonzeros---close to two orders of magnitude---that their heuristic achieves in two-player poker endgames.
That is important due to the central role of poker in imperfect-information game solving.
Poker variants have been the standard canonical benchmarks in game theory since the introduction of the most seminal solution concept, Nash equilibrium, in 1950~\citep{Nash50:Equilibrium,Kuhn50:Simplified}. Poker captures the essence of private information and strategic, game-theoretic deception and reasoning. In fact, in Nash's dissertation, the only application was poker~\citep{Nash50:Non}. In the ensuing decades, larger and larger poker variants were tackled in AI~\citep{Waterman70:Generalization} and operations research~\citep{Zadeh77:Computation}. Then around year 2000, poker was recognized as the main challenge problem for imperfect-information game solving in AI~\citep{Billings02:Challenge}. Hundreds of papers have been published on it, the AAAI Annual Computer Poker Competition was organized, and superhuman AI performance has been achieved~\cite{Bowling15:Heads,Brown17:Superhuman,Brown19:Superhuman}. This has dramatically pushed the boundary of imperfect-information game-solving technology. As many questions in the field remain open (for example, the computation of interpretable, sparse, collusive, or sequentially-rational strategies), we are convinced that poker will continue to play a fundamental role as the gold standard in imperfect-information games for decades to come.

We show that the existence of extremely sparse factorizations in poker games can be tied to their particular \emph{Kronecker-product} structure. The existence of that structure was mentioned  by  \citet{Hoda10:Smoothing} solely for the purpose of reducing memory footprint of their first-order method for Nash equilibrium, and \citet{Johanson11:Accelerating} use essentially the same structure (without the Kroenecker representation) to speed up best-response computation in poker. We clarify how Kroenecker-product structure arises and, most importantly, introduce the connection between that structure and sparsification. By leveraging the Kronecker-product structure directly, we give two ways of computing strong sparsifications of poker games (as well as any other game with a similar structure). We show that our sparsification techniques are i) orders of magnitude faster to compute, ii) more numerically stable, and iii) produce a dramatically smaller number of nonzeros than the general algorithm by \citet{Zhang20:Sparsified}. Our sparsification techniques enable---for the first time---effective computation of high-precision Nash equilibria and strategies subject to constraints on the amount of allowed randomization. Furthermore, they significantly speed up parallel first-order game-solving algorithms; we show state-of-the-art speed on GPU.

\paragraph{Weaknesses}
While our techniques apply to all games with a Kronecker-product structure (that is, whose payoff matrix can be expressed as a sum of Kronecker products), currently the only games with practical relevance that are known to exhibit a Kronecker-product structure are poker games. 
That said, as many questions in the field remain open (for example, the computation of interpretable, sparse, collusive, or sequentially-rational strategies), we believe that poker will continue to play a fundamental role as the gold standard in imperfect-information games for decades to come.

As we show, our techniques have the concrete potential to help make a dent on those important questions by enabling one to scale up existing optimization methods---essentially for free---by replacing the payoff matrix of the game with its sparsified counterpart.
\section{Payoff matrix sparsification and its applications}\label{sec:applications}

Extensive-form games are played on a game tree and can capture both sequential and simultaneous moves, stochastic events (such as a roll of the dice, or drawing a random card from a shuffled deck) as well as private information. A strategy for a generic Player~$i$ in an extensive-form game is an assignment of probability to each of the player's \emph{sequences}---that is, sequence of actions that the player can take starting from the root of the game tree. Just like in normal-form games, the outcomes of a two-player extensive-form game can be arranged compactly into a \emph{payoff matrix} $\vec{A}$, whose rows and columns are indexed over all sequences of the two players. Specifically, let $z$ be an outcome (terminal state) of the game tree, let $u$ be the payoff assigned to Player~$i$ by that outcome, and let $\sigma_i,\sigma_j$ be the sequences for Player~$i$ and her opponent, respectively, corresponding to $z$. Finally, let $c$ be the product of the probability of all stochastic events on the path from the root of the game tree to $z$. Then, Player~$i$'s payoff matrix contains, on the row corresponding to sequence $\sigma_i$ and column corresponding to $\sigma_j$, a payoff equal to $u\cdot c$.
For the purposes of this paper, a \emph{sparsification} of the payoff matrix
$\vec{A}$ of a game will be defined as an expression of the form
\begin{equation}\label{eq:sparsification}
    \vec{A} = \hat{\vec{A}} + \vec{U}\vec{M}^{-1}\vec{V}^\top,
\end{equation}
for suitable matrices $\hat{\vec{A}}, \vec{U}, \vec{M}$ and $\vec{V}$, where
$\vec{M}$ is an invertible triangular matrix. The expression in~\eqref{eq:sparsification}
is more general than the one considered by \citet{Zhang20:Sparsified}, which
corresponds to the case where $\vec{M}$ is the identity matrix. We
will show in \cref{sec:experiments} how the flexibility afforded by the matrix
$\vec{M}$  translates into better performance. Given a sparsification of
$\vec{A}$, we will refer to its \emph{size} as the sum of the number of
nonzeros of the matrices $\hat{\vec{A}}, \vec{U}, \vec{V}$ and $\vec{M}$. A ``good'' sparsification is one whose size is significantly
smaller than the number of nonzeros of the original matrix $\vec{A}$.
We will investigate three main applications of payoff matrix sparsification.

\paragraph{1. Linear programming and high-precision Nash equilibrium strategies.}
It is well-known that a Nash equilibrium strategy for a player in a two-player zero-sum perfect-recall
extensive-form game can be expressed as the solution to an LP by using the \emph{sequence-form representation} \citep{Stengel96:Efficient,Koller96:Efficient,Romanovskii62:Reduction}.
Specifically, given the payoff matrix $\vec{A}$ (say, for Player 1), a Nash equilibrium
strategy for that player is the solution to the LP on the left
of~\eqref{eq:lp}, where the matrices $\vec{F}_{i}$ and vectors $\vec{f}_{i}$ (for $i \in \{1,2\}$)
are very sparse and define the sequence-form constraints for Player 1 and 2, respectively.
\begin{figure}
    \begin{equation}\label{eq:lp}\small
        \arraycolsep=1.0pt
        \mleft\lbrace\begin{array}{l}
            \displaystyle\max~ \vec{f}_2^\top \vec{v} \\[2mm]
            \arraycolsep=1.4pt
            \begin{array}[t]{ll}
                \circled{1} & \vec{A}^\top \vec{x} - \vec{F}_2^\top \vec{v} \ge \vec{0} \\[1mm]
                \circled{2} & \vec{F}_1\vec{x}                      = \vec{f}_1         \\[1mm]
                \circled{3} & \vec{x} \ge \vec{0}, \vec{v}~~\text{free}
            \end{array} 
        \end{array}\mright.
        \to
        \mleft\lbrace\begin{array}{l}
            \displaystyle\max~ \vec{f}_2^\top \vec{v} \\[2mm]
            \arraycolsep=1.4pt
            \begin{array}[t]{ll}
                \circled{1} & \hat{\vec{A}}^\top \vec{x} - \vec{F}_2^\top \vec{v} + \vec{V}\vec{w} \ge \vec{0} \\[1mm]
                \circled{2} & \vec{F}_1\vec{x}                      = \vec{f}_1                                \\[1mm]
                \circled{3} & \vec{U}^\top \vec{x} -\vec{M}^\top \vec{w} =\vec{0}                              \\[1mm]
                \circled{4} & \vec{x} \ge \vec{0}, \vec{v}~~\text{free}, \vec{w}~~\text{free}.
            \end{array}
        \end{array}\mright.
    \end{equation}
\end{figure}
With a sparsification of $\vec{A}$, the LP on the left of \eqref{eq:lp}
can be rewritten as the one on the right, trading the number of nonzeros of $\vec{A}$ for the size of the sparsification on the right.
When the size of the sparsification is much smaller than the number of nonzeros of $\vec{A}$,
the LP on the right is significantly sparser, and can therefore be
solved much faster (or at all, in large games) by LP technology. 
That enables the computation of
Nash equilibrium strategies at a high level of precision, a task that is infeasible for
iterative first-order methods (such as CFR \citep{Zinkevich07:Regret} and EGT \citep{Hoda10:Smoothing,Kroer18:Solving}). One immediate application of computing
Nash equilibrium strategies at that level of precision is the ability to
compute the \emph{exact} value of the game. Another important reason is that the computation of optimal, basic strategies (that is, vertices of the LP) represents a
fundamental building block in the computation of sequentially-rational
equilibrium refinements~\citep{Farina18:Practical}.
In \cref{sec:applications} we show that our sparsification techniques enable one to compute high-precision Nash equilibrium strategies in games significantly larger than what was possible with the sparsification technique of~\citet{Zhang20:Sparsified}.

\paragraph{2. Integer programming and least-exploitabile deterministic strategies.}
Deterministic strategies can be deployed without
the need for randomization---at which humans are notoriously bad---and
are arguably more interpretable than randomized strategies. How
much randomization is needed to play optimally in poker is a long-standing open question. (Some early work on simplified models has suggested that not much randomization is needed~\citep{Chen06:Mathematics,Ganzfried10:Computing}.)
Our sparsification techniques help scale the computation of strategies
subject to constraints on the amount of required randomization. For instance,
a least-exploitable deterministic strategy can be computed as the
solution to the integer program obtained from either formulation in~\eqref{eq:lp}
by replacing the constraint $\vec{x} \ge \vec{0}$ with the constraints that
$\vec{x}$ be a vector of binary variables. In large games, even state-of-the-art commercial integer programming
technology cannot even remotely scale up to the size of the unsparsified formulation. Instead, in \cref{sec:experiments} we will show that
the same formulation sparsified with our techniques enables---to our knowledge, for the first time---the computation of provably near-optimal deterministic strategies.
%
We will also measure how much less value a deterministic player can guarantee herself---a metric we coin \emph{price of determinism}.
We find that in the real no-limit Texas hold'em endgames we test on, the price of determinism is minimal:
deterministic strategies extract at least $98.26\%$ of the
value of the game in all cases. Given the benefits of deterministic strategies
(such as lower memory requirement, no need to randomize, higher interpretabilty, and ease of deployment by humans),
we believe this to be an interesting positive experimental outcome on a long-standing research question that also
warrants further investigation. 

\paragraph{3. First-order methods and highly-parallel gradient computation.}
First-order methods that compute approximate Nash equilibrium
strategies---such as CFR~\citep{Zinkevich07:Regret} and EGT~\citep{Hoda10:Smoothing,Kroer18:Solving}---require, as an intermediate step at each iteration, the
evaluation of the gradient of the utility function, which can be
computed via a sparse matrix-vector multiplication between the
payoff matrix $\vec{A}$ and the strategy $\vec{x}$ of a player.
Given a sparsification of $\vec{A}$, the following is a natural algorithm for computing
$\vec{A}\vec{x}$: first, compute the product
$\vec{y} \defeq \vec{V}^\top\vec{x}$; then, solve the sparse
triangular system $\vec{M}\vec{z} = \vec{y}$, solving for $\vec{z}$ (we skip this step when $\vec{M}$ is the identity matrix, and instead immediately let $\vec{z} = \vec{y}$);
then, multiply the solution $\vec{z}$ of the triangular system by $\vec{U}$, computing $\vec{w}\defeq \vec{U}\vec{z}$; finally, sum
the sparse matrix-vector product $\hat{\vec{A}}\vec{x}$ to $\vec{w}$.
Each of the matrix-vector products involved requires a number of operations proportional to the number of nonzeros of the matrix. Furthermore, since $\vec{M}$ is triangular, $\vec{z}$ can be computed in time proportional to the number of nonzeros in $\vec{M}$. So, the number of floating-point operations required by the algorithm is proportional to the
size of the sparsification. 
Given the wide availability of highly-tuned libraries for sparse matrix-vector
multiplication both for CPUs and GPUs, the method we have just described
enables an extremely concise and efficient implementation of the gradient of
the utility function of sparsified games, which can easily rival specialized combinatorial algorithms~\citep{Johanson11:Accelerating}.
%

\section{Kronecker-product structure of poker games}\label{sec:kronecker structure}

In this section, we illustrate and formalize a particular combinatorial
structure---which we refer to as \emph{Kronecker-product structure}---that
poker games possess. Only a basic working knowledge of poker is needed to follow this section. In the
appendix we describe the basic rules of poker.
The term \emph{Kronecker-product structure} refers to the fact that the payoff
matrix can be expressed as (a sum of) terms of the form
\begin{equation*}\small
    \vec{P}\kro\vec{Q} \defeq \begin{bmatrix} P_{11}\vec{Q} & \cdots & P_{1n}\vec{Q} \\
                \vdots        & \ddots & \vdots        \\
                P_{m1}\vec{Q} & \cdots & P_{mn}\vec{Q}
    \end{bmatrix} \in \bbR^{(mr)\times(ns)},
\end{equation*}
for appropriate matrices $\vec{P}\in\bbR^{m\times n}$ and $\vec{Q}\in\bbR^{r\times s}$ and arbitrary dimensions $m,n,r,s$.
%

In this section, we shed  light on how this structure arises,
%
by focusing
on the endgame that begins immediately after the last (aka. river) card is revealed---called a \emph{river endgame}. The
observations we will make about river endgames in this subsection apply
more generally to the endgame that begins immediately after the turn and the
flop, as well as the full poker game. We will conventionally refer to the first mover in
the endgame (that is, the ``small blind'' player) as `Player~$1$', and to the
second mover (the ``big blind'' player) as `Player~$2$'. We will focus on computing
the payoff matrix for Player~$1$; the payoff matrix for Player~$2$ is completely analogous.

\begin{figure*}[th]
    \centering\scalebox{.99}{%
        \begin{tikzpicture}[xscale=1,>=latex']
            \def\done{.6*1.6}
            \def\dtwo{.4*1.6}
            \def\dvert{1.1}
            \def\acsz{8.75}
            \def\acpos{.567}
            \def\sqpos{.3498}
            \tikzstyle{edg}=[semithick];
            \tikzstyle{acsty}=[fill=white,inner xsep=.3mm,inner ysep=.5mm];
            \tikzstyle{sqsty}=[];
            \tikzstyle{nodelbl}=[black,scale=.9,black!90!white];

            \node[text width=3.1cm,anchor=north west] at (-7.5,0.4) {
                {\small Community cards:}\\[1.2mm]
                \includegraphics[scale=.7]{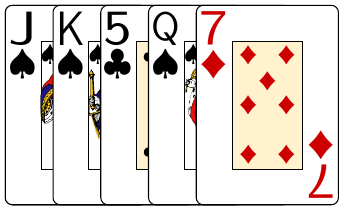}
            };

            \node[draw,text width=2.705cm,anchor=north west,fill=black!10!white] at (6.95,0.4) {
                \small Pot contributions (\$)
            };
            \node[draw,text width=2.705cm,anchor=north west] at (6.95,-0.12) {
            \!{\setlength{\tabcolsep}{1.8mm}%
            \fontsize{7.7}{7.9}\selectfont\scalebox{.9}{\begin{tabular}{c|rr}
                    \textsc{f} & 1,875.0  & 1,875.0  \\         
                    \textsc{h} & 20,000.0 & 1,875.0  \\        
                    \textsc{i} & 20,000.0 & 20,000.0 \\       
                    \textsc{j} & 4,687.5  & 4,687.5  \\     
                    \textsc{l} & 4,687.5  & 1,875.0  \\       
                    \textsc{n} & 4,687.5  & 4,687.5  \\     
                    \textsc{p} & 1,875.0  & 4,687.5  \\       
                    \textsc{r} & 1,875.0  & 20,000.0 \\        
                    \textsc{s} & 20,000.0 & 20,000.0 \\       
                    \textsc{t} & 11,718.8 & 11,718.8 \\   
                    \textsc{v} & 4,687.5  & 11,718.8 \\    
                    \textsc{w} & 4,687.5  & 20,000.0 \\      
                    \textsc{x} & 20,000.0 & 20,000.0 \\       
                    \textsc{y} & 11,718.8 & 4,687.5  \\  
                    \textsc{z} & 11,718.8 & 11,718.8 \\   
                    $\Phi$     & 20,000.0 & 4,687.5  \\          
                    $\Theta$   & 20,000.0 & 20,000.0 \\         
                    $\Lambda$  & 20,000.0 & 11,718.8 \\
                    $\Xi$      & 20,000.0 & 20,000.0 \\
                    $\Psi$     & 11,718.8 & 20,000.0 \\
                    $\Omega$   & 20,000.0 & 20,000.0 \\
                \end{tabular}}}
            };

            \node[pl1] (X0) at (0,0) {}; \node[left=1.123mm of X0,nodelbl] {\textsc{a}};
            \node[pl2] (Y1) at ($(X0) + (-3, -\dvert)$) {}; \node[left=1.123mm of Y1,nodelbl] {\textsc{b}};
            \node[pl2] (Y30) at ($(X0) + (0, -\dvert)$) {}; \node[right=1.123mm of Y30,nodelbl] {\textsc{c}};
            \node[pl2] (Y18) at ($(X0) + (3, -\dvert)$) {}; \node[right=1.123mm of Y18,nodelbl] {\textsc{d}};
            \node[pl1] (X3) at ($(Y1) + (-1.2, -\dvert)$) {}; \node[left=1.123mm of X3,nodelbl] {\textsc{e}};
            \node[showdown] (Z2) at ($(Y1) + (0.0, -\dvert)$) {}; \node[left=0.4789mm of Z2,nodelbl] {\textsc{f}};
            \node[pl1] (X15) at ($(Y1) + (1.2, -\dvert)$) {};\node[left=1.123mm of X15,nodelbl] {\textsc{g}};
            \node[fold] (Z31) at ($(Y30) + (-.5, -\dvert)$) {};\node[left=0.4789mm of Z31,nodelbl] {\textsc{h}};
            \node[showdown] (Z32) at ($(Y30) + (.5, -\dvert)$) {};\node[left=0.4789mm of Z32,nodelbl] {\textsc{i}};
            \node[showdown] (Z20) at ($(Y18) + (-1.5, -\dvert)$) {};\node[left=0.4789mm of Z20,nodelbl] {\textsc{j}};
            \node[pl1] (X21) at ($(Y18) + (-.25, -\dvert)$) {};\node[left=1.123mm of X21,nodelbl] {\textsc{k}};
            \node[fold] (Z19) at ($(Y18) + (1, -\dvert)$) {};\node[right=.8mm of Z19,nodelbl] {\textsc{l}};
            \node[pl1] (X27) at ($(Y18) + (2.25, -\dvert)$) {};\node[right=1.123mm of X27,nodelbl] {\textsc{m}};
            \node[showdown] (Z5) at ($(X3) + (-2, -\dvert)$) {};\node[below=.5mm of Z5,nodelbl] {\textsc{n}};
            \node[pl2] (Y6) at ($(X3) + (-.8333, -\dvert)$) {};\node[right=1.123mm of Y6,nodelbl] {\textsc{o}};
            \node[fold] (Z4) at ($(X3) + (0.3333, -\dvert)$) {};\node[below=.5mm of Z4,nodelbl] {\textsc{p}};
            \node[pl2] (Y12) at ($(X3) + (1.5, -\dvert)$) {};\node[left=1.123mm of Y12,nodelbl] {\textsc{q}};
            \node[fold] (Z16) at ($(X15) + (0, -\dvert)$) {};\node[right=0.4789mm of Z16,nodelbl] {\textsc{r}};
            \node[showdown] (Z17) at ($(X15) + (1, -\dvert)$) {};\node[right=0.4789mm of Z17,nodelbl] {\textsc{s}};
            \node[fold] (Z7) at ($(Y6) + (-1, -\dvert)$) {}; \node[below=.5mm of Z7,nodelbl] {\textsc{y}};
            \node[showdown] (Z8) at ($(Y6) + (0, -\dvert)$) {}; \node[below=.5mm of Z8,nodelbl] {\textsc{z}};
            \node[pl1] (X9) at ($(Y6) + (1, -\dvert)$) {};\node[left=1.123mm of X9,nodelbl] {$\Gamma$};
            \node[fold] (Z13) at ($(Y12) + (-.5, -\dvert)$) {};\node[right=0.4789mm of Z13,nodelbl] {$\Phi$};
            \node[showdown] (Z14) at ($(Y12) + (.5, -\dvert)$) {};\node[right=0.4789mm of Z14,nodelbl] {$\Theta$};
            \node[fold] (Z10) at ($(X9) + (-.5, -\dvert)$) {};\node[left=0.4789mm of Z10,nodelbl] {$\Psi$};
            \node[showdown] (Z11) at ($(X9) + (.5, -\dvert)$) {};\node[right=0.4789mm of Z11,nodelbl] {$\Omega$};
            \node[showdown] (Z23) at ($(X21) + (-1.2, -\dvert)$) {};\node[left=0.4789mm of Z23,nodelbl] {\textsc{t}};
            \node[pl2] (Y24) at ($(X21) + (0, -\dvert)$) {}; \node[left=1.123mm of Y24,nodelbl] {\textsc{u}};
            \node[fold] (Z22) at ($(X21) + (1.2, -\dvert)$) {};\node[left=0.4789mm of Z22,nodelbl] {\textsc{v}};
            \node[fold] (Z28) at ($(X27) + (-.5, -\dvert)$) {};\node[below=.5mm of Z28,nodelbl] {\textsc{w}};
            \node[showdown] (Z29) at ($(X27) + (.5, -\dvert)$) {};  \node[below=.5mm of Z29,nodelbl] {\textsc{x}};
            \node[fold] (Z25) at ($(Y24) + (-.5, -\dvert)$) {};\node[right=0.4789mm of Z25,nodelbl] {$\Lambda$};
            \node[showdown] (Z26) at ($(Y24) + (.5, -\dvert)$) {};\node[right=0.4789mm of Z26,nodelbl] {$\Xi$};

            \draw [edg,->] (X0) -- (Y1) node[pos=.7,acsty] {\fontsize{\acsz}{\acsz}\selectfont check} node[pos=\sqpos,sqsty]{\small\hideseq{1}};
            \draw [edg,->] (X0) -- (Y30) node[pos=.6,acsty] {\fontsize{\acsz}{\acsz}\selectfont all~in} node[pos=\sqpos,sqsty]{\small\hideseq{2}};
            \draw [edg,->] (X0) -- (Y18) node[pos=.7,acsty] {\fontsize{\acsz}{\acsz}\selectfont bet} node[pos=\sqpos,sqsty]{\small\hideseq{3}};
            \draw [edg,dashed,->] (Y1) -- (X3) node[pos=\acpos,acsty] {\fontsize{\acsz}{\acsz}\selectfont bet} node[pos=\sqpos,sqsty]{\small\hideseq{1}};
            \draw [edg,dashed,->] (Y1) -- (Z2) node[pos=\acpos,acsty] {\fontsize{\acsz}{\acsz}\selectfont check} node[pos=\sqpos,sqsty]{\small\hideseq{2}};
            \draw [edg,dashed,->] (Y1) -- (X15) node[pos=\acpos,acsty,xshift=1mm] {\fontsize{\acsz}{\acsz}\selectfont all~in} node[pos=\sqpos,sqsty]{\small\hideseq{3}};
            \draw [edg,dashed,->] (Y30) -- (Z31)  node[pos=\acpos,acsty] {\fontsize{\acsz}{\acsz}\selectfont fold} node[pos=\sqpos,sqsty]{\small\hideseq{9}};
            \draw [edg,dashed,->] (Y30) -- (Z32) node[pos=\acpos,acsty] {\fontsize{\acsz}{\acsz}\selectfont call} node[pos=\sqpos,sqsty]{\small\hideseq{10}};
            \draw [edg,dashed,->] (Y18) -- (Z20) node[pos=\acpos,acsty] {\fontsize{\acsz}{\acsz}\selectfont call} node[pos=\sqpos,sqsty]{\small\hideseq{11}};
            \draw [edg,dashed,->] (Y18) -- (X21) node[pos=\acpos,acsty] {\fontsize{\acsz}{\acsz}\selectfont raise} node[pos=\sqpos,sqsty]{\small\hideseq{12}};
            \draw [edg,dashed,->] (Y18) -- (Z19) node[pos=\acpos,acsty] {\fontsize{\acsz}{\acsz}\selectfont fold} node[pos=\sqpos,sqsty]{\small\hideseq{13}};
            \draw [edg,dashed,->] (Y18) -- (X27) node[pos=\acpos,acsty] {\fontsize{\acsz}{\acsz}\selectfont all~in} node[pos=\sqpos,sqsty]{\small\hideseq{14}};
            \draw [edg,->] (X3) -- (Z5) node[pos=\acpos,acsty] {\fontsize{\acsz}{\acsz}\selectfont call} node[pos=\sqpos,sqsty]{\small\hideseq{4}};
            \draw [edg,->] (X3) -- (Y6) node[pos=\acpos,acsty] {\fontsize{\acsz}{\acsz}\selectfont raise} node[pos=\sqpos,sqsty]{\small\hideseq{5}};
            \draw [edg,->] (X3) -- (Z4) node[pos=\acpos,acsty] {\fontsize{\acsz}{\acsz}\selectfont fold} node[pos=\sqpos,sqsty]{\small\hideseq{6}};
            \draw [edg,->] (X3) -- (Y12) node[pos=\acpos,acsty] {\fontsize{\acsz}{\acsz}\selectfont all~in} node[pos=\sqpos,sqsty]{\small\hideseq{7}};
            \draw [edg,->] (X15) -- (Z16) node[pos=\acpos,acsty] {\fontsize{\acsz}{\acsz}\selectfont fold} node[pos=\sqpos,sqsty,xshift=-.3mm]{\small\hideseq{10}};
            \draw [edg,->] (X15) -- (Z17) node[pos=\acpos,acsty] {\fontsize{\acsz}{\acsz}\selectfont call} node[pos=\sqpos,sqsty,xshift=.3mm]{\small\hideseq{11}};
            \draw [edg,dashed,->] (Y6) -- (Z7) node[pos=\acpos,acsty] {\fontsize{\acsz}{\acsz}\selectfont fold} node[pos=\sqpos,sqsty]{\small\hideseq{4}};
            \draw [edg,dashed,->] (Y6) -- (Z8) node[pos=\acpos,acsty] {\fontsize{\acsz}{\acsz}\selectfont call} node[pos=\sqpos,sqsty]{\small\hideseq{5}};
            \draw [edg,dashed,->] (Y6) -- (X9)node[pos=\acpos,acsty] {\fontsize{\acsz}{\acsz}\selectfont all~in} node[pos=\sqpos,sqsty]{\small\hideseq{6}};
            \draw [edg,dashed,->] (Y12) -- (Z13) node[pos=\acpos,acsty] {\fontsize{\acsz}{\acsz}\selectfont fold} node[pos=\sqpos,sqsty]{\small\hideseq{7}};
            \draw [edg,dashed,->] (Y12) -- (Z14) node[pos=\acpos,acsty] {\fontsize{\acsz}{\acsz}\selectfont call} node[pos=\sqpos,sqsty]{\small\hideseq{8}};
            \draw [edg,->] (X9) -- (Z10) node[pos=\acpos,acsty] {\fontsize{\acsz}{\acsz}\selectfont fold} node[pos=\sqpos,sqsty]{\small\hideseq{8}};
            \draw [edg,->] (X9) -- (Z11) node[pos=\acpos,acsty] {\fontsize{\acsz}{\acsz}\selectfont call} node[pos=\sqpos,sqsty]{\small\hideseq{9}};
            \draw [edg,->] (X21) -- (Z23) node[pos=\acpos,acsty] {\fontsize{\acsz}{\acsz}\selectfont call} node[pos=\sqpos,sqsty]{\small\hideseq{12}};
            \draw [edg,->] (X21) -- (Y24) node[pos=\acpos,acsty] {\fontsize{\acsz}{\acsz}\selectfont all~in} node[pos=\sqpos,sqsty]{\small\hideseq{13}};
            \draw [edg,->] (X21) -- (Z22) node[pos=\acpos,acsty] {\fontsize{\acsz}{\acsz}\selectfont fold} node[pos=\sqpos,sqsty]{\small\hideseq{14}};
            \draw [edg,->] (X27) -- (Z28)node[pos=\acpos,acsty] {\fontsize{\acsz}{\acsz}\selectfont fold} node[pos=\sqpos,sqsty,xshift=-.3mm]{\small\hideseq{15}};
            \draw [edg,->] (X27) -- (Z29)  node[pos=\acpos,acsty] {\fontsize{\acsz}{\acsz}\selectfont call} node[pos=\sqpos,sqsty,xshift=.3mm]{\small\hideseq{16}};
            \draw [edg,dashed,->] (Y24) -- (Z25) node[pos=\acpos,acsty] {\fontsize{\acsz}{\acsz}\selectfont fold} node[pos=\sqpos,sqsty,xshift=-.3mm]{\small\hideseq{15}};
            \draw [edg,dashed,->] (Y24) -- (Z26) node[pos=\acpos,acsty] {\fontsize{\acsz}{\acsz}\selectfont call} node[pos=\sqpos,sqsty,xshift=.3mm]{\small\hideseq{16}};
        \end{tikzpicture}}\vspace{-2mm}
    \caption{Skeleton of a river endgame. The initial pot contributions for the endgame are $\$1875$ for both players. Each player has a stack size worth $\$18125$ they can play. Every `bet' and `raise' action corresponds to first matching the other player's contribution to the pot, and then increasing the player's contribution to the pot by an amount equal to $\nicefrac{3}{4}$ of the cumulative amount in the pot.
        Black nodes belong to the small blind player, white nodes to the big blind player. The symbol \protect\tikz{\protect\node[showdown] at (0,0) {};} denotes a showdown, while \protect\tikz{\protect\node[fold] at (0,0) {};} denotes that one player folded. 
    }\vspace{-2mm}
    \label{fig:river skeleton}
\end{figure*}

In a river endgame, all community cards have already been revealed, and
the two players engage in a single round of betting before the endgame
ends. To fully describe a particular instance of a river endgame, the following
quantities must be given:
(i) The collection $B$ of five community cards (the \emph{board}) that have been drawn;
(ii) Initial stack sizes $(s_1, s_2)$, the amount of money that Player~$1$ and~$2$, respectively, possess in their stack at the beginning of the endgame;
(iii) Initial pot contribution $c$, the amount of money that have been contributed to the pot by Player~$1$ and~$2$, prior to the endgame;
(iv) Two belief distributions $\mu_i: \mathcal{H}_i \to [0,1]$, one for each player $i\in\{1,2\}$, assigning a probability distribution to each possible hand of the players.%
\footnote{Usually, the belief distributions reflect the posterior that each player has over the hands of the opponent, given what they have observed about the opponent's play prior to the river endgame. 
    Here, we make no assumption on how the belief distributions have been formed, and simply take the two distributions as given.}
The river endgame is an extensive-form game of its own,
where at the root of the game tree a chance node assigns private hands $(h_1,
    h_2)\in \mathcal{H}_1\times\mathcal{H}_2$ compatible with $B$ (that is, so that
when putting together the hands and the board, no card appears more than once)
to each player according to the distribution
\[\small
    \pi(h_1,h_2) = \frac{1}{\beta}\begin{cases}
        0                    & \text{if } h_1, h_2, B \text{ are incompatible} \\
        \mu_1(h_1)\mu_2(h_2) & \text{otherwise},
    \end{cases}
\]
where $\beta$ is the appropriate normalization constant so that $\sum_{h_1,h_2}\pi(h_1,h_2) = 1$.
Then, the game proceeds with one betting round (with the standard mechanics recalled in \cref{sec:poker}), which can either end with a player folding, or with a showdown.
The actions that the players can take in the betting round is the same,
\emph{regardless of their private hands}. In other words, the subtrees
rooted under each possible outcome of the root chance nodes (which
corresponds to an assignment of hands for each player), are all equal. To study the combinatorial properties of the game tree
corresponding to the river endgame, it
is then natural to only focus on \emph{one}, generic
such subtree, which we call the \emph{skeleton} of the
river endgame.
\cref{fig:river skeleton}  depicts the skeleton of a river
endgame for a very coarse betting abstraction.%
%
%
%
The payoff matrix $\vec{A}$ for any player can be expressed as a block matrix $\left[\vec{A}_{h_1,h_2} \right]_{h_1\in\mathcal{H}_1,h_2\in\mathcal{H}_2} $, where each block $\vec{A}_{h_1, h_2}$ is the matrix arising from playing the skeleton when the hands of the players are set to $h_1$ and $h_2$, rescaled by the probability of the pair of hands, $\pi(h_1,h_2)$ specified earlier.
The main goal of this section is to show that as the pair of hands $(h_1,h_2)$ varies, the blocks $\vec{A}_{h_1,h_2}$ exhibit very little variability. That regular structure will then enable us to express the payoff matrix $\vec{A}$ of Player~$1$ as a sum of Kronecker products of suitable matrices.

Fix any pair of hands $\left( h_1, h_2
    \right)\in\mathcal{H}_1\times\mathcal{H}_2$. The block
$\vec{A}_{h_1,h_2}$ of the river endgame's payoff matrix for Player~$1$ tabulates the payoffs
corresponding to the terminal states that can be reached when the players
are dealt hands
$h_1,h_2$ (a stochastic events that occurs with probability
$\pi(h_1,h_2)$, as defined above). Since the mechanics of the betting round do not depend on the choice of hands, those terminal states are exactly the same terminal states that can be reached in the skeleton of the river endgame. So, the block $\vec{A}_{h_1, h_2}$ can be written as
$
    \vec{A}_{h_1,h_2} = \pi(h_1,h_2) \vec{A}^\text{skel}_{h_1,h_2},
$
where $\vec{A}^\text{skel}_{h_1,h_2}$ is the payoff matrix induced by the skeleton when the players' hands are set to $h_1$ and $h_2$. Furthermore, by separating the contributions $\vec{F}$ and $\vec{S}_{h_1,
        h_2}$ from fold and showdown terminal states respectively, the payoff matrix
$\vec{A}^\text{skel}_{h_1,h_2}$ can be written as
$
    \vec{A}^\text{skel}_{h_1,h_2} = \vec{F} + \vec{S}_{h_1,h_2}.
$
The matrix $\vec{F}$ of payoffs associated to the fold terminal states is
straightforward to compute. The initial stacks and pot contributions are known,
so the stacks and pot contributions of the players at each node of the skeleton
can be easily determined by following the path of (betting) actions from the
root of the skeleton to that state (see \cref{fig:river skeleton} (Right), and the appendix for a worked out example). 
%
We now turn our attention to the matrix of showdown payoffs $\vec{S}_{h_1,h_2}$.
When the Player~$1$'s hand beats the opponent's, the payoff of the player at each
showdown terminal state is equal to Player~$2$'s pot contribution---which, by the rules of poker, is equal to
Player~$1$'s pot contribution. When the
player's hand loses the opponent's, the payoff at each showdown terminal state
is the negative amount of the player's pot contribution. Finally, when the
hands tie (or are incompatible given the board), the payoffs are all zero. So,
introducing the quantity $\gamma(h_1,h_2)$ defined as $1$ when hand
$h_1$ beats $h_2$ (given the board $B$ that was dealt), $-1$ when hand $h_1$ is beaten by hand $h_2$, and $0$
when the hands tie or are incompatible given the board, we can write $\vec{S}_{h_1, h_2} = \gamma(h_1, h_2)\vec{S}$,
where $\vec{S}$ is the matrix of Player~$1$'s pot contributions at each of showdown
terminal states of the skeleton. So, putting all the observations together, we have that
$\vec{A}_{h_1,h_2} = \pi(h_1,h_2) \vec{F} + \pi(h_1, h_2) \gamma(h_1,h_2) \vec{S}$ for
all hand pairs $(h_1, h_2)\in\mathcal{H}_1\times\mathcal{H}_2$, and we are
ready to formalize the Kronecker-product structure of river endgames in formal terms.

\begin{proposition}\label{pr:river}
    Consider a river endgame with board $B$ and hand beliefs $\mu_1,\mu_2$ with normalization constant $\beta$, and let $\vec{F}$ and $\vec{S}$
    be the matrices of fold payoffs and showdown payoffs as described above. Introduce the vectors $\vec{\lambda}_{i}$ and diagonal matrices $\vec{\Lambda}_{i}$ for each player $i=1,2$, whose entries are indexed over hands and are defined as
    $
        \vec{\lambda}_i[h_i] = \vec{\Lambda}_i[h_i,h_i] \defeq \frac{\mu_i\left( h_i \right) }{\sqrt{\beta}} \quad\forall i\in\{1,2\}, h_i \in \mathcal{H}_i.
    $
    Furthermore, introduce the $|\mathcal{H}_1|\times|\mathcal{H}_2|$ matrices $\vec{H}^\times, \vec{W},$ and $\vec{C}$, defined as
    \begin{align*}
        \begin{array}{c}
            \vec{W}[h_1,h_2]\defeq\gamma(h_1,h_2),\quad \vec{C} \defeq \vec{\mu}_1\vec{\mu}_2^\top - \vec{\Lambda}_1\vec{H}^\times \vec{\Lambda}_2,
            \\[2mm]
            \vec{H}^\times[h_1,h_2]\defeq\begin{cases}
                1 & \begin{minipage}{5cm}if $h_1,h_2$, and $B$ are incompatible\end{minipage} \\
                0 & \text{otherwise}.
            \end{cases}
        \end{array}
    \end{align*}
    Then, the payoff matrix $\vec{A}$ for Player~$1$ can be written as the sum of
    Kronecker products
    \begin{equation}\label{eq:A}
        \vec{A} = \vec{C}   \kro \vec{F} + (\vec{\Lambda}_1\vec{W}\vec{\Lambda}_2) \kro \vec{S}.
    \end{equation}
\end{proposition}

The ideas presented so far were presented in the context of a river endgame,
but they apply directly also to the endgame that starts right after the turn card
has been revealed, and more broadly in the full game tree of poker.
For the turn endgame, we would start from the skeleton of the first betting round. Only
two outcomes are possible: either the game ends in a fold, or the betting round terminates
in a non-fold terminal state $z$, at which point the final card (aka. river card) is revealed and a river
endgame begins. Note that because the river card is public, the payoff matrix of
the turn endgame is made of diagonal blocks, with each block representing a river endgame. Because of
the diagonal structure, each river endgame can be independently decomposed as in \cref{pr:river}
and sparsified using the techniques we will develop in the next section. This line of reasoning
can be composed for each of betting rounds in the game. That shows that \cref{pr:river} in
fact captures the essence of the combinatorial, Kronecker-structure nature of poker games.

\section{Sparsification techniques}\label{sec:technique}

We propose two sparsification techniques that directly leverage the Kronecker-product structure of the
payoff matrix that we described in \cref{sec:kronecker structure}. We will do
so with reference to the same symbols used in \cref{pr:river}.  
We will find the following property of the Kronecker product useful.
%
\begin{property}[Mixed-product rule]\label{pro:mixed product}
    Let $\vec{P}\in\bbR^{m\times n},\vec{Q}\in\bbR^{r\times s},\vec{C}\in\bbR^{n\times \ell},\vec{D}\in\bbR^{s\times q}$ be arbitrary matrices. Then, $(\vec{P}\vec{C})\kro(\vec{Q}\vec{D}) = (\vec{P}\kro\vec{Q})(\vec{C}\kro\vec{D}).$
\end{property}
The two techniques we propose operate on expression~\eqref{eq:A} by sparsifying
its two terms $\vec{C}\kro\vec{F}$
and $(\vec{\Lambda}_1\vec{W}\vec{\Lambda}_2)\kro\vec{S}$ separately by
fundamentally using the mixed-product rule for Kronecker products. Both techniques sparsify the term $\vec{C}\kro\vec{F}$ using the same
strategy:
\begin{align}
    &\hspace{-2mm}\vec{C}\kro \vec{F}
    = (\vec{\mu}_1\,\vec{\mu}_2^{\!\top} - \vec{\Lambda}_1\vec{H}^\times\vec{\Lambda}_2)\kro \vec{F}\nonumber\\
    &= -(\vec{\Lambda}_1\vec{H}^\times\vec{\Lambda}_2) \kro \vec{F} + (\vec{\mu}_1\,\vec{\mu}_2^{\!\top})\kro (\vec{I} \vec{F}) \nonumber\\
                        &= -(\vec{\Lambda}_1\vec{H}^\times\vec{\Lambda}_2) \kro \vec{F} + (\vec{\mu}_1 \kro \vec{I}) (\vec{\mu}_2 \kro \vec{F}^\top)^\top,\label{eq:cf}
\end{align}
where we used the bilinearity of Kronecker products in the second equality, and the mixed product rule in the last one. The two techniques differ in the way they handle the term $(\vec{\Lambda}_1\vec{W}\vec{\Lambda}_2)\kro\vec{S}$ in~\eqref{eq:A}.

\xsubsection{Technique A}\label{sec:technA}
The first technique sparsifies the term $(\vec{\Lambda}_1\vec{W}\vec{\Lambda}_2)\kro\vec{S}$ by 
recursively sparsifying the `win-lose' matrix $\vec{W}$.
Specifically, it first computes a
sparsification $\vec{W} = \hat{\vec{W}} + \vec{U}_W\vec{V}_W^\top$ (in our experiments,
we do so by using the general heuristic described in \citep{Zhang20:Sparsified}),
and then uses the mixed-product rule of Kronecker product to write (all steps are in the appendix)
\begin{align}
    &(\vec{\Lambda}_1\vec{W}\vec{\Lambda}_2)\kro \vec{S} 
                                                          = (\vec{\Lambda}_1\hat{\vec{W}}\vec{\Lambda}_2) \kro \vec{S}\nonumber\\ &\hspace{1cm}+ \Big((\vec{\Lambda}_1\vec{U}_W) \kro \vec{I}\Big) \Big((\vec{\Lambda}_2\vec{V}_W) \kro \vec{S}^\top\Big)^{\!\top},\label{eq:technA}
\end{align}
where 
 the equality follows from the mixed-product rule.
Putting \eqref{eq:cf} and \eqref{eq:technA} together, we obtain: 
\begin{proposition}
The payoff matrix \eqref{eq:A} admits the sparsification $\vec{A} = \hat{\vec{A}} + \vec{U}\vec{M}^{-1}\vec{V}^\top$, where
\[
\begin{array}{c}
    \hat{\vec{A}} \defeq (\vec{\Lambda}_1\hat{\vec{W}}\vec{\Lambda}_2)\kro \vec{S} - (\vec{\Lambda}_1\vec{H}^\times\vec{\Lambda}_2) \kro \vec{F},\\[2mm]
    \vec{U} \defeq \Big[(\vec{\Lambda}_1\vec{U}_W) \kro \vec{I}~\Big|~ \vec{\mu}_1\kro \vec{I}\Big],\qquad \vec{M} \defeq \vec{I},\\[2mm]
    \vec{V} \defeq \Big[(\vec{\Lambda}_2\vec{V}_W) \kro \vec{S}^\top ~\Big|~ \vec{\mu}_2\kro \vec{F}^\top\Big].
\end{array}
\]
\end{proposition}

\xsubsection{Technique B}\label{sec:technB}
The second technique leverages the fact that the hands of each player can be ranked
by their strength. When that is done (ignore for now incompatible hands) each row of the win-lose matrix $\vec{W}$ begins with zero or more
columns equal to $-1$, followed by zero or more columns with value $0$, followed by zero or more columns with
value $1$. As the hand of the row player becomes stronger, the number of $-1$'s on the row decreases, and the number of $1$'s increases. Hence, the matrix obtained by subtracting from each row of $\vec{W}$ the previous line must be very sparse. We can
compactly represent the operation of subtracting from each row of $\vec{W}$ the preceding row via the matrix operation $\vec{Y} \defeq \vec{D}\vec{W}$,
where the lower bidiagonal matrix $\vec{D}$ has value $1$ on the main diagonal, and value $-1$ in the diagonal below the main diagonal. Then,
\begin{align}
    &(\vec{\Lambda}_1\vec{W}\vec{\Lambda}_2)\kro \vec{S} = \Big(\vec{\Lambda}_1(\vec{D}^{-1} \vec{Y})\vec{\Lambda}_2\Big)\kro (\vec{I}\vec{S})
                                                       \nonumber \\&\hspace{1cm}= (\vec{\Lambda}_1\kro\vec{I})(\vec{D}\kro\vec{I})^{-1}\Big((\vec{\Lambda}_2\vec{Y}^\top) \kro \vec{S}^\top\Big)^{\!\top}\!\!,\label{eq:technB}
\end{align}
and we can state the following result.
\begin{proposition}
The payoff matrix \eqref{eq:A} admits the sparsification $\vec{A} = \hat{\vec{A}} + \vec{U}\vec{M}^{-1}\vec{V}^\top$, where
\[
\begin{array}{c}
    \hat{\vec{A}} \defeq - (\vec{\Lambda}_1\vec{H}^\times\vec{\Lambda}_2) \kro \vec{F},\\[1mm]
    \vec{U} \defeq \Big[\vec{\Lambda}_1 \kro \vec{I}~\Big|~ \vec{\mu}_1\kro \vec{I}\Big],\qquad
    \vec{M} \defeq 
    \renewcommand\arraystretch{1.3}
    \mleft[
    \begin{array}{c|c}
      \vec{D}\kro\vec{I} & \\
      \hline
      & \vec{I}
    \end{array}
    \mright],\\[3mm]
    \vec{V} \defeq \Big[(\vec{\Lambda}_2\vec{Y}^\top) \kro \vec{S}^\top ~\Big|~ \vec{\mu}_2\kro \vec{F}^\top\Big].
\end{array}
\]
\end{proposition}

\xsubsection{Postprocessing}\label{sec:postprocessing}
After computing any payoff matrix sparsification, we further slightly decrease its size by removing columns from $\vec{V}$ that are identically zero. This process is perhaps best exemplified in the case of Technique A, where $\vec{M} = \vec{I}$. Suppose that the $j$-th column of $\vec{V}$ is zero. Then, given any vector $\vec{x}$, the $j$-th row of the vector $\vec{V}^\top\vec{x}$ will be zero. Hence, we can safely discard the $j$-th column of $\vec{U}$, potentially decreasing the size of the sparsification. 

When $\vec{M}$ is not the identity (as is the case for Technique B), the process is only slightly more involved. In the rest of the discussion, we will assume that $\vec{M}$ is lower triangular, and that all the entries on its main diagonal are equal to $1$. Suppose that the $j$-th column of $\vec{V}$ is identically zero. Then, the $j$-th row of $\vec{V}^\top\vec{x}$ is zero, for any vector $\vec{x}$. We can take advantage of that fact when computing $\vec{M}^{-1}\vec{V}^\top\vec{x}$, that is, when solving the system $\vec{M}\vec{y} = \vec{V}^\top\vec{X}$. In particular, the $j$-th row of the system is of the form $\vec{y}_j + \sum_{i < j} a_i \vec{y}_i = 0$, which implies that $\vec{y}_j = -\sum_{i<j}a_i\vec{y}_i$. Hence, the $j$-th entry of $\vec{y}$ is a linear combination of other rows of $\vec{y}$ and does not need to be stored explicitly. In other words, we can remove any reference to $\vec{y}_j$ from the system, and replace it with $-\sum_{i<j} a_i\vec{y}_i$. In the case of Technique B, that operation is especially cheap, given that each row of $\vec{M}$ always has at most two nonzeros (so, $\vec{y}_j$ is simply substituted with $\vec{y}_i$ for some $i < j$). Because $\vec{y}_j$ is treated implicitly as a linear combination of other entries in $\vec{y} = \vec{M}^{-1}\vec{V}^\top\vec{x}$ we can simply adjust $\vec{U}$ by removing the $j$-th column from the matrix, and sum it, multiplied by $a_i$, to column $i$.
\section{Experimental results}
\label{sec:experiments}

We experimentally compare the sparsification techniques
introduced in \cref{sec:technique} on eight River endgames that were actually played in the
\emph{Brains vs AI} competition where superhuman performance was reached by an AI, \textit{Libratus}, against four top specialist professional players in no-limit Texas hold'em in January 2017. 
Unless otherwise indicated, each endgame uses
the betting abstraction used by \textit{Libratus} (a description is available in the
appendix), which contains significantly more bet sizes than the simple betting
abstraction used in the sparsification experiments of \citet{Zhang20:Sparsified}; so, we are addressing significantly larger games. In all games, isomorphic hands~\citep{Gilpin07:Lossless,Johanson11:Accelerating,Waugh13:Fast}
were collapsed into a single meta-hand, as is standard in computational experiments on poker. All
experiments were conducted on a computer with 32GB of RAM and an Intel
CPU with 16 (virtual) cores, each with a nominal speed of 2.40GHz.

\xsubsection{Computing the sparsification}\label{sec:exp sparsif}
We compare the sparsifications computed by the two techniques we introduced in
\cref{sec:technique} against the general sparsification technique of~\citet{Zhang20:Sparsified}. We compare both the time to compute the
sparsification and the size (\emph{i.e.}, number of nonzeros) of the resulting sparsification.  We ran the
iterative algorithm of \citet{Zhang20:Sparsified} with a fixed random seed and a cap on the number of
sparsifying iterations set to $1000$. We reused the same implementation of
the general technique of \citeauthor{Zhang20:Sparsified} in the implementation of our
Technique A to sparsify matrix $\vec{W}$. We
implemented all algorithms in C++, using the Eigen library to provide the
implementation of linear algebraic objects such as sparse matrices and vectors.\footnote{\citet{Zhang20:Sparsified} recommend using a custom implementation for implicit matrices to enhance performance. Judging from the results in their paper, that modification would not change our evaluation. For example, when using a small betting abstraction, they report that their optimized implementation took 68s seconds on river endgame 7 (the smallest game we test on). Our techniques, on the significantly larger betting abstraction we test on, took less than 500ms for the same game.}
Full results are in available in \cref{tab:sparsification time}.

\begin{table*}\centering
    \centering
    \newcommand{\snum}[1]{%
        \num[scientific-notation=true,
            round-mode=figures,
            round-precision=3]{#1}}
    \newcommand{\oom}{\textcolor{gray}{oom}}
    \scalebox{.93}{\begin{tikzpicture}
            \node[anchor=west] at (0,0) {
                \begin{tabular}{lr|||rr|||rr|||rr}
                    \toprule
                    \multirow{2}{*}{\textbf{Game}} & \multirow{2}{*}{\textbf{Unsparsified size}}
                                                   & \multicolumn{2}{c|||}                                                                                                              
                    {\textbf{Zhang \& Sandholm}}
                                                   & \multicolumn{2}{c|||}{\textbf{Technique A}}
                                                   & \multicolumn{2}{c}{\textbf{Technique B}}
                    \\
                                                   &                                             & Size            & Time     & Size           & Time      & Size           & Time      \\
                    \midrule
                    River 7                        & \snum{50925699}                             & \snum{1739420}  & \n{751}  & \snum{406660}  & \n{0.42}  & \snum{274046}  & \n{0.318} \\
                    River 6                        & \snum{60293211}                             & \snum{1965786}  & \n{900}  & \snum{363704}  & \n{0.58}  & \snum{269956}  & \n{0.454} \\
                    River 8                        & \snum{95876527}                             & \snum{4148753}  & \n{2052} & \snum{562560}  & \n{0.594} & \snum{392580}  & \n{0.436} \\
                    River 2                        & \snum{177243650}                            & \snum{11012151} & \n{8517} & \snum{1015630} & \n{0.748} & \snum{672028}  & \n{0.567} \\
                    River 4                        & \snum{221113174}                            & \snum{11038177} & \n{9006} & \snum{1259286} & \n{0.48}  & \snum{775760}  & \n{0.624} \\
                    River 1                        & \snum{447454764}                            & \oom            & \oom     & \snum{2269957} & \n{0.889} & \snum{1604276} & \n{0.699} \\
                    River 3                        & \snum{475580270}                            & \oom            & \oom     & \snum{2756858} & \n{1.035} & \snum{1645453} & \n{0.722} \\
                    River 5                        & \snum{479162564}                            & \oom            & \oom     & \snum{2521310} & \n{1.016} & \snum{1646190} & \n{0.733} \\
                    \bottomrule
                \end{tabular}
            };
            \draw[line width=2.2mm,white] (4.61,-2.2) -- +(0,4.4);
            \draw[line width=2.2mm,white] (8.41,-2.2) -- +(0,4.4);
            \draw[line width=2.2mm,white] (11.93,-2.2) -- +(0,4.4);
        \end{tikzpicture}}
    \caption{Comparison between different sparsification techniques. 
        `oom': out of memory.}
    \label{tab:sparsification time}
\end{table*}

The algorithm by \citeauthor{Zhang20:Sparsified} could scale up to river
endgame 4 (a game with 220 million terminal states) before running out of
memory. Our techniques could handle all eight endgames. In terms of sparsity,
the technique by \citeauthor{Zhang20:Sparsified} is able to consistently reduce
the number of nonzeros required to represent the payoff matrix by a factor in
the range 20-50. Our Technique A increases sparsity by a
factor between 100 and 200. Our Technique B increases sparsity
by a factor between 200 and 400, producing sparsifications that are consistently
roughly twice as small as Technique A. In terms of time required by the
sparsification algorithm to compute the sparsification in memory, the algorithm
by \citeauthor{Zhang20:Sparsified} requires an amount of time in the order of
hours, whereas our techniques require between 300 milliseconds and 1 second to
compute the sparsification by directly leveraging the Kronecker structure of
the endgame. In summary, our techniques consistently produce
dramatically better sparsifications while at the same time requiring orders of
magnitude less compute time to generate.

\xsubsection{Computation of an optimal basis for Nash equilibrium}
In this subsection we show that our sparsification techniques enable---to our
knowledge for the first time in the large endgames we test on---the computation of a
Nash equilibrium strategy that is an optimal basic (i.e., vertex) solution to the Nash equilibrium LP~\eqref{eq:lp}, as discussed in the first bullet point
of \cref{sec:applications}.
In our experiments we used the
state-of-the-art solver Gurobi to solve the LP. Full results can be found in
\cref{tab:exact nash time}, where we measures the time required by
Gurobi to solve the LPs, not including the time required to
compute the sparsifications (where applicable).

In all games, we solved for a
strategy for Player 1.
When the LP was left unsparsified, the solver could barely start,
immediately running out of memory in River 8. We avoided running experiments with
the unsparsified LP beyond River 8. The technique of \citet{Zhang20:Sparsified} (set up as described in~\cref{sec:exp sparsif})
did not run out of memory, but caused Gurobi to terminate abnormally due to
numeric instability in River 8 and River 2. In the games for which the unsparsified
LP and the LP sparsified using Zhang and Sandholm's technique
could be solved, the performance of Gurobi on the latter was 5x-60x worse than with our sparsification techniques. Using our techniques, we were able to compute---for the first time---an optimal basis for Nash equilibrium (and correspondigly, the exact
value of the game) in all eight river endgames. Overall, Technique B outperformed Technique A
in the larger games by a margin of 1x--3x.

\begin{table}[H]
    \centering
    \newcommand{\snum}[1]{%
        \num[scientific-notation=true,
            round-mode=figures,
            round-precision=3]{#1}}
    \newcommand{\oom}{\textcolor{gray}{oom}}
    \newcommand{\nt}{\textcolor{red}{\sffamily{\scriptsize NT}}}
    \newcommand{\numeric}{\textcolor{gray}{trouble}}
    \newcommand{\dnr}{\textcolor{gray}{---}}
    \newcommand{\win}{$\star$}
    \scalebox{.93}{\begin{tabular}{lrrrr}
            \toprule
            \textbf{Game}
                    & \textbf{Unsparsif.}
                    & \textbf{ZS20}                                                             
                    & \textbf{Techn. A}
                    & \textbf{Techn. B}
            \\
            \midrule
            River 7 & $8$m $51$s          & $2$m $38$s & \win{} \n{24.14}  & \n{27.09}          \\
            River 6 & $2$m $35$s          & $6$m $07$s & \win{} \n{6.83}   & \n{7.29}           \\
            River 8 & \oom                & \numeric   & \win{} $1$m $55$s & $2$m $43$s         \\
            River 2 & \dnr                & \numeric   & $38$m $34$s       & \win{} $21$m $8$s  \\
            River 4 & \dnr                & \dnr       & $21$m $55$s       & \win{} $17$m $18$s \\
            River 1 & \dnr                & \dnr       & $2$h $58$m        & \win{} $2$h $18$m  \\
            River 3 & \dnr                & \dnr       & $3$h $54$m        & \win{} $3$h $17$m  \\
            River 5 & \dnr                & \dnr       & $7$h $09$m        & \win{} $2$h $34$m  \\
            \bottomrule
        \end{tabular}}
    \caption{Computation of an optimal basis for Nash equilibrium.
        `oom': out of memory. `trouble': Gurobi indicated a numeric error in its log.}
    \label{tab:exact nash time}
\end{table}

\xsubsection{Computation of a least-exploitable deterministic strategy}
In this subsection we investigate another application that is enabled for the
first time by our sparsification technique: the computation of the
least-exploitable (that is, strongest against a fully rational agent,
aka. minimax) \textit{deterministic} strategy, as described in the second bullet point of
\cref{sec:applications}.
This application relies on the ability to run
linear integer programming technology; in our experiments we used the state-of-the-art solver Gurobi.
We investigate computing least-exploitable deterministic strategies in all eight
river endgames, using the full \textit{Libratus} betting abstraction in the three smallest
games and endgame 4, and the smaller betting abstraction used by \citet{Zhang20:Sparsified} in
the remaining games as Gurobi struggled to solve the larger games with the larger
abstraction. We only report data for games sparsified with Technique B, which was found
to be the most scalable technique in the previous subsection. In \cref{tab:pod} we
report, for each endgame, an upper bound on the price of determinism that Gurobi was able
to certify, and how long it took Gurobi to reach that price of determinism. The experiments show that
deterministic strategies are able to extract at least $\approx 98\%$ of the value of in all the games!

\begin{table}[H]
    \centering
    \newcommand{\snum}[1]{%
        \num[scientific-notation=true,
            round-mode=figures,
            round-precision=3]{#1}}
    \newcommand{\oom}{\textcolor{gray}{oom}}
    \newcommand{\nt}{\textcolor{red}{\sffamily{\scriptsize NT}}}
    \newcommand{\numeric}{\textcolor{gray}{trouble}}
    \newcommand{\dnr}{\textcolor{gray}{---}}
    \newcommand{\win}{$\star$}
    \scalebox{.93}{\begin{tabular}{lrrrr}
            \toprule
            \textbf{Game}
                                      & \textbf{Time}
                                      & \textbf{Price of determ.}
            \\
            \midrule
            River 7                   & \n{581}                   & $< 1.16\%$ \\
            River 6                   & \n{1529}                  & $< 1.18\%$ \\
            River 8                   & \n{708}                   & $< 1.10\%$ \\
            River 2\smash{$^\dagger$} & \n{4286}                  & $< 1.74\%$ \\
            River 4                   & \n{61749}                 & $< 1.42\%$ \\
            River 1\smash{$^\dagger$} & \n{2103}                  & $< 1.00\%$ \\
            River 3\smash{$^\dagger$} & \n{2437}                  & $< 0.94\%$ \\
            River 5\smash{$^\dagger$} & \n{3795}                  & $< 1.60\%$ \\
            \bottomrule
        \end{tabular}}
    \caption{Computation of a least-exploita\-ble deterministic strategy for Player 1, and the corresponding price of determinism.
    }
    \label{tab:pod}
\end{table}
\daggerfootnote{The experiments marked with this symbol were conducted on the smaller betting abstraction of \citet{Zhang20:Sparsified} rather than the original one used by \textit{Libratus}, because Gurobi was far from a good solution after 12 hours on the larger betting abstraction.}

\xsubsection{First-order methods}
\begin{figure*}[t]\centering%
    \begin{minipage}[b]{\textwidth}\centering
        \definecolor{kcol}{HTML}{377eb8}
        \definecolor{mcol}{HTML}{4daf4a}
        \definecolor{briancol}{HTML}{f781bf}
        \definecolor{xdensecol}{HTML}{ff7f00}
        \definecolor{noamcol}{HTML}{800080}
        \includegraphics[scale=.78]{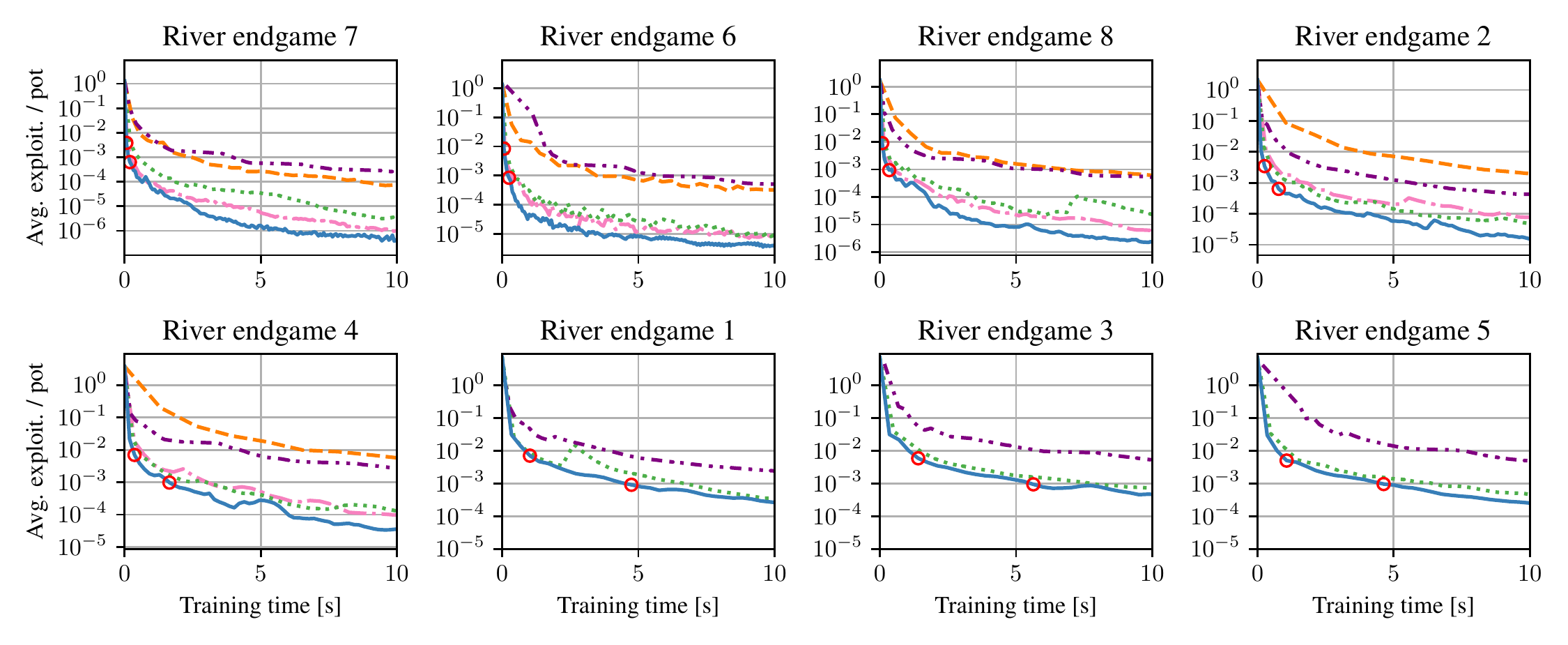}\\[-1mm]
        \scalebox{1}{\begin{tikzpicture}
                \draw[kcol,very thick] (-.2,0) -- (.3,0);
                \node[anchor=west] at (.3,0) {\small GPU (A)};
                \draw[dotted,mcol,very thick] (2.0,0) -- (2.5,0);
                \node[anchor=west] at (2.5,0) {\small GPU (B)};
                \draw[dashdotted,briancol,very thick] (4.2,0) -- (4.7,0);
                \node[anchor=west] at (4.7,0) {\small GPU (ZS20)};
                \draw[dashed,xdensecol,very thick] (6.8,0) -- (7.3,0);
                \node[anchor=west] at (7.3,0) {\small GPU (Unsparsified)};
                \draw[noamcol,very thick] (10.4,0)--(10.56666,0) (10.677777,0)--(10.733333,0) (10.84444,0)--(10.9,0);
                \node[anchor=west] at (10.9,0) {\small CPU poker-specific};
                \draw[gray] (-.4,-.27) rectangle (13.7,.27);
            \end{tikzpicture}}
        \caption{Approximate Nash equilibrium computation using different first-order
            methods, including our GPU implementation of Discounted CFR leveraging the
            sparsification techniques. The red circles mark the first time `GPU (A)' reaches average exploitability less than $1\%$ and $0.1\%$.}
        \label{fig:gpu}
    \end{minipage}
\end{figure*}
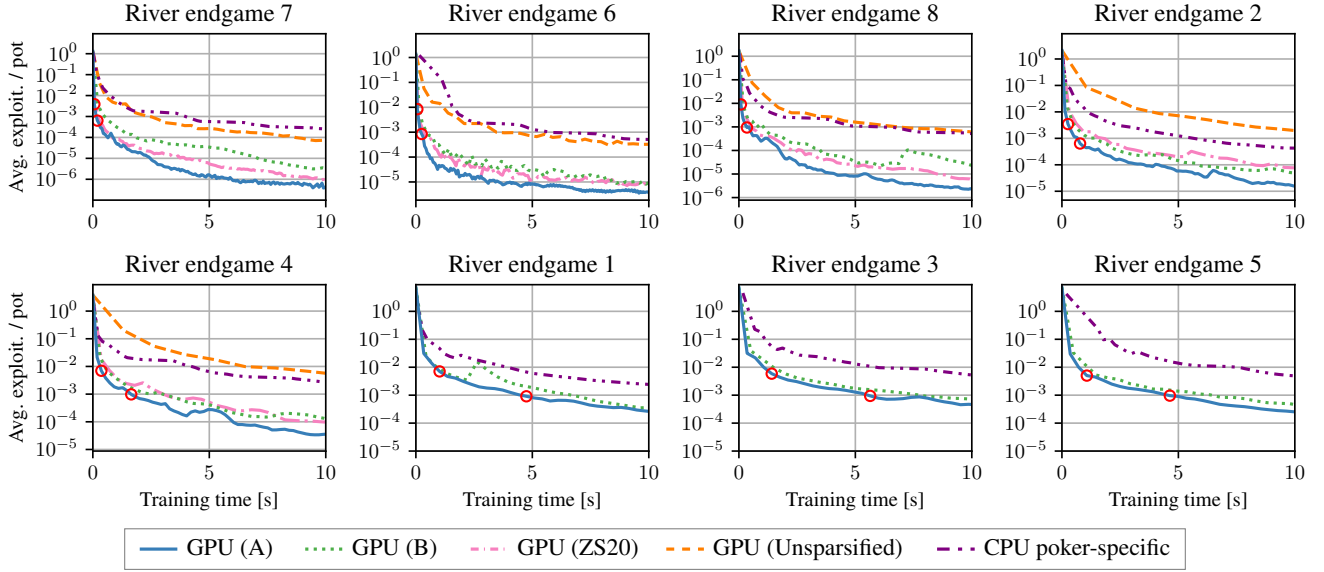

As mentioned in the third bullet point of \cref{sec:applications}, our sparsification
techniques enable a straightforward parallel method for computing the gradients
of the utility function of the game required by first-order methods at each iterations to
compute approximate Nash equilibrium strategies.
To showcase that application, we implemented a GPU version of the state-of-the-art CFR variant for poker, \textit{Discounted CFR (DCFR)}~\citep{Brown19:Solving}.
Our GPU version of the algorithm was implemented
within Nvidia's
CUDA framework and run on a laptop-grade Quadro T2000 GPU. It gains from parallelism by updating strategies in parallel for each
possible hands of the players. We use the highly-tuned Cusparse libraries to
represent, manipulate, and operate on, sparse matrices. We compare two versions
of the same code.
The first, which we call `\emph{GPU (Unspars.)}' in \cref{fig:gpu},
computes each gradient $\vec{A}\vec{x}$ by explicitly performing the
matrix-vector multiplication.
The second version leverages the payoff matrix sparsification
$\vec{A} = \hat{\vec{A}} + \vec{U}\vec{M}^{-1}\vec{V}^\top$ to compute the gradient $\vec{A}\vec{x}$ as
described in the third bullet of \cref{sec:applications}.
Depending on which sparsification is used, we call this version
of the GPU implementation `\emph{GPU (ZS20)}' (for \citet{Zhang20:Sparsified}), `\emph{GPU (A)}' and `\emph{GPU (B)}'.
%
We also compared against a parallel, CPU-based state-of-the-art poker-specific
implementation of DCFR that includes the computational shortcuts described by
\citet{Johanson11:Accelerating}, denoted `\emph{CPU poker-specific}'.
That algorithm had access to all 16 CPU cores. Results are
in \cref{fig:gpu}. The y axis measures the average exploitability of
the strategy profile within the betting abstraction (equal to half of the Nash saddle point gap), normalized by the total amount of money in the pot at the
beginning of the river endgame. Strategies with a relative exploitability of
$0.1\%$-$1\%$ are generally considered suitable for play against top human poker
professionals. The x axis measures wall-clock time, not including the
time to compute the sparsification of the payoff matrix (where
applicable). Our Technique A consistently
outperforms Technique B, due the absence of the extra operation of solving a
triangular system. Our GPU implementation based on Technique A
significantly outperforms all other algorithms, and is able to compute strong
approximate Nash equilibrium strategies suitable for play against human poker
professionals within $5$-$6$ seconds in the worst case, including the time
required to compute the sparsification. In many games, it required less than
two seconds. These times are well within the norms of usual speed of poker play.

\section{Conclusions}

We showed that \emph{Kronecker structure} present in games enables
the design of specialized payoff matrix sparsification techniques. Those 
techniques in turn enable optimization algorithms (such as interior-point methods, the simplex method, and integer programming technology) to scale to real-world poker endgames
that were previously impossible to handle for those methods, due to the huge size
of the payoff matrix of the game. The ability to apply out-of-the-box commercial
solvers in games is important, as it enables one to quickly explore questions
such as the computation of exact (within numeric tolerance) Nash equilibria, vertex Nash equilibria needed for equilibrium refinements, and least-exploitable deterministic strategies. Furthermore, they significantly speed up parallel first-order game-solving algorithms. We show state-of-the-art speed on a GPU.


\section*{Acknowledgments}
This material is based on work supported by the National Science Foundation under grants IIS-1718457, IIS-1901403, and CCF-1733556, and the ARO under award W911NF2010081.
We thank the anonymous referees for useful comments, and Noam Brown for providing the poker endgames and CPU-based state-of-the-art poker-specific
implementation of DCFR that we use in the experiments.

\bibliography{dairefs}


\iftrue
    \onecolumn
    \clearpage
    \appendix
    \section{Poker games}\label{sec:poker}

Many variants of poker exist. The results in this paper apply to many of them,
including limit hold'em poker, no-limit hold'em poker, and Rhode Island poker.
Furthermore, the results of this note apply to \emph{endgames} of those poker
variants, which are attractive benchmark games in the computational game
solving literature due to their more tractable size.

For ease of treatment, all examples in this note refer to no-limit hold'em
(NLHE) poker---the most popular variant of poker in the world. In the interest
of keeping these notes self-contained, in the rest of this section we recall
the rules of NLHE in the measure that they are relevant for understanding our
discussion and results in the rest of the paper. Irrelevant aspects, such as
tie-breaking rules in the determination of the relative strength of poker
hands, will be ignored.

At all times in a two-player NLHE game, each of the two players is associated
with three quantities: (1) a private \emph{hand}, consisting of two cards; (2)
a public \emph{stack}, representing the (nonnegative) amount of money that the
player has at his or her disposal; and (3) a public \emph{pot contribution
  amount}, corresponding to the total amount of money that the player has moved
from their stack to a shared \emph{pot}. Furthermore, one of the players is
designated as the \emph{big blind player}, and the other player as the
\emph{small blind player}.
At the beginning of the game, the big blind player and the small blind player
move fixed amounts of money (called the \emph{big blind} and \emph{small
  blind}, respectively) from their stack to the pot. Then, the two players are
dealt their private hand, which are drawn from a standard 52-card shuffled
deck. At that point, one \emph{betting round} start. In the betting round, the
players take one or more turns, in which they have an option to increase their
stake in the game (that is, their contributions to the pot) as they see fit,
according to rules which will be detailed later, or to bail out from the
game---which is commonly known as \emph{folding} the player's hand. If a player
folds, the game ends and the player that has \emph{not} folded takes the amount
in the pot. If by the end of the betting round no player has folded, three
cards---called the \emph{flop}---are dealt publicly (face-up) from the deck.
Those revealed cards are observed by both player, and for that reason are
called \emph{community cards}.
After the flop is dealt, another betting round starts. If by the end of the
second betting round no player has folded their hands, a single community
card---called the \emph{turn}---is revealed. After that, a third betting round
begins. If no player folds in that third betting round, a final community
card---the \emph{river}---is dealt, and one last betting round starts. If no
fold occurs in that last betting round, a \emph{showdown} occurs. In a
showdown, the hands of the players are revealed, and the player with the
strongest hand (determined according to criteria mentioned later in this
section) takes all the money in the pot and moves it onto their own stack. If
the hands of the players are equally strong, the pot is split evenly by the two
players. The payoff of each player is the difference between their stack amount
at the end of the game and at the beginning of the game.

\paragraph{Betting rounds} In each betting round, the two players have a chance
to increase their stake in the game, by moving money from their stack to the
shared pot, or to \emph{fold} their hand, thereby terminating the game and
letting the opponent take all of the money in the pot. Each betting round
proceeds in turns, where the two players alternate at each turn.
In the first betting round (before the flop is revealed), the first player to
act in the betting round is the big blind. In all other betting rounds, the
first player to act is the small blind.

At any given turn, the actions that are available to the acting player depend
on whether or not the contribution of that player to the pot is strictly lower
than the contribution of the opponent, or not.
\begin{itemize}[nolistsep,itemsep=1mm,leftmargin=5mm]
  \item If the acting player's contribution is not strictly lower than the opponent's, the player can either \emph{check}---that is, not move any money from their stack to the pot---or \emph{bet} any positive amount of money in their stack, by moving that amount from the stack into the pot.
  \item If the acting player's contribution is strictly lower than the opponent's, the player has an option to either i) \emph{fold} their hand, ii) \emph{call} the previous bet (move money from the stack to the pot until the contributions are equal, or the player has finished the money), or iii) \emph{raise} on the previous bet, that is, first calling the previous bet and then moving any desired positive amount of remaining money from the player's stack into the pot, up to at most as much money as the opponent has in their stack.
\end{itemize}
If a player decides to contribute the maximum amount in they are allowed into the pot, the player is said to have gone \emph{all-in}.
The betting round ends when either player folds, one player calls, or two players check.

\paragraph{Betting abstraction}
As stated above, in NLHE players can bet any amount of money they possess in their stack. However, in order to make poker tractable computationally, existing methods discretize the possible amounts into a finite set of standard bet amounts, called a \emph{betting abstraction}. Those bet amounts are usually expressed as a fraction of the pot. In this paper, we assume that a discretization has been fixed, but otherwise make no assumption on what discretization has been set, and our results apply to every betting abstraction that might have been chosen.

\paragraph{Showdowns}
If no player ever folds, the game ends with a \textit{showdown}. In that case, the hands of the players get revealed. Each player identifies the strongest five-card poker hand they can form by combining their two own cards together with the five face-up community cards.
The details as to how the relative strength of a hand over another is determined are beyond the scope of this paper, and can be easily found online or in books. We simply note that given the set of five community cards, the hands of a player can be sorted from weakest to strongest, according to a total order.

\section{Further examples around the game of \cref{fig:river skeleton}}
\begin{figure*}[th]
  \centering\scalebox{.9}{%
    \begin{tikzpicture}[xscale=1,>=latex']
      \def\done{.6*1.6}
      \def\dtwo{.4*1.6}
      \def\dvert{1.35}
      \def\acsz{8.75}
      \def\acpos{.667}
      \def\sqpos{.3498}
      \tikzstyle{edg}=[semithick];
      \tikzstyle{acsty}=[fill=white,inner xsep=.3mm,inner ysep=.5mm];
      \tikzstyle{sqsty}=[fill=black!10!white,scale=.85,inner xsep=.3mm,inner ysep=.5mm,text=black!80!white];
      \tikzstyle{nodelbl}=[black,scale=.9,black!90!white];

      \node[text width=3.1cm,anchor=north west] at (-7.5,0.4) {
        {\small Community cards:}\\[1.2mm]
        \includegraphics[scale=.7]{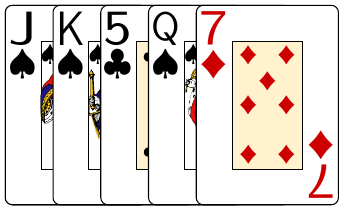}
      };

      \node[draw,text width=3cm,anchor=north west,fill=black!10!white] at (6.95,0.4) {
        \small Pot contributions (\$)
      };
      \node[draw,text width=3cm,anchor=north west] at (6.95,-0.12) {
      \!{\setlength{\tabcolsep}{1.8mm}%
      \small\scalebox{.9}{\begin{tabular}{c|rr}
          \textsc{f} & 1875    & 1875    \\         
          \textsc{h} & 20000   & 1875    \\        
          \textsc{i} & 20000   & 20000   \\       
          \textsc{j} & 4687.5  & 4687.5  \\     
          \textsc{l} & 4687.5  & 1875    \\       
          \textsc{n} & 4687.5  & 4687.5  \\     
          \textsc{p} & 1875    & 4687.5  \\       
          \textsc{r} & 1875    & 20000   \\        
          \textsc{s} & 20000   & 20000   \\       
          \textsc{t} & 11718.8 & 11718.8 \\   
          \textsc{v} & 4687.5  & 11718.8 \\    
          \textsc{w} & 4687.5  & 20000   \\      
          \textsc{x} & 20000   & 20000   \\       
          \textsc{y} & 11718.8 & 4687.5  \\  
          \textsc{z} & 11718.8 & 11718.8 \\   
          $\Phi$     & 20000   & 4687.5  \\          
          $\Theta$   & 20000   & 20000   \\         
          $\Lambda$  & 20000   & 11718.8 \\
          $\Xi$      & 20000   & 20000   \\
          $\Psi$     & 11718.8 & 20000   \\
          $\Omega$   & 20000   & 20000   \\
        \end{tabular}}}
      };

      \node[pl1] (X0) at (0,0) {}; \node[left=1.123mm of X0,nodelbl] {\textsc{a}};
      \node[pl2] (Y1) at ($(X0) + (-3, -\dvert)$) {}; \node[left=1.123mm of Y1,nodelbl] {\textsc{b}};
      \node[pl2] (Y30) at ($(X0) + (0, -\dvert)$) {}; \node[right=1.123mm of Y30,nodelbl] {\textsc{c}};
      \node[pl2] (Y18) at ($(X0) + (3, -\dvert)$) {}; \node[right=1.123mm of Y18,nodelbl] {\textsc{d}};
      \node[pl1] (X3) at ($(Y1) + (-1.2, -\dvert)$) {}; \node[left=1.123mm of X3,nodelbl] {\textsc{e}};
      \node[showdown] (Z2) at ($(Y1) + (0.0, -\dvert)$) {}; \node[left=0.4789mm of Z2,nodelbl] {\textsc{f}};
      \node[pl1] (X15) at ($(Y1) + (1.2, -\dvert)$) {};\node[left=1.123mm of X15,nodelbl] {\textsc{g}};
      \node[fold] (Z31) at ($(Y30) + (-.5, -\dvert)$) {};\node[left=0.4789mm of Z31,nodelbl] {\textsc{h}};
      \node[showdown] (Z32) at ($(Y30) + (.5, -\dvert)$) {};\node[left=0.4789mm of Z32,nodelbl] {\textsc{i}};
      \node[showdown] (Z20) at ($(Y18) + (-1.5, -\dvert)$) {};\node[left=0.4789mm of Z20,nodelbl] {\textsc{j}};
      \node[pl1] (X21) at ($(Y18) + (-.25, -\dvert)$) {};\node[left=1.123mm of X21,nodelbl] {\textsc{k}};
      \node[fold] (Z19) at ($(Y18) + (1, -\dvert)$) {};\node[right=.8mm of Z19,nodelbl] {\textsc{l}};
      \node[pl1] (X27) at ($(Y18) + (2.25, -\dvert)$) {};\node[right=1.123mm of X27,nodelbl] {\textsc{m}};
      \node[showdown] (Z5) at ($(X3) + (-2, -\dvert)$) {};\node[below=.5mm of Z5,nodelbl] {\textsc{n}};
      \node[pl2] (Y6) at ($(X3) + (-.8333, -\dvert)$) {};\node[right=1.123mm of Y6,nodelbl] {\textsc{o}};
      \node[fold] (Z4) at ($(X3) + (0.3333, -\dvert)$) {};\node[below=.5mm of Z4,nodelbl] {\textsc{p}};
      \node[pl2] (Y12) at ($(X3) + (1.5, -\dvert)$) {};\node[left=1.123mm of Y12,nodelbl] {\textsc{q}};
      \node[fold] (Z16) at ($(X15) + (0, -\dvert)$) {};\node[right=0.4789mm of Z16,nodelbl] {\textsc{r}};
      \node[showdown] (Z17) at ($(X15) + (1, -\dvert)$) {};\node[right=0.4789mm of Z17,nodelbl] {\textsc{s}};
      \node[fold] (Z7) at ($(Y6) + (-1, -\dvert)$) {}; \node[below=.5mm of Z7,nodelbl] {\textsc{y}};
      \node[showdown] (Z8) at ($(Y6) + (0, -\dvert)$) {}; \node[below=.5mm of Z8,nodelbl] {\textsc{z}};
      \node[pl1] (X9) at ($(Y6) + (1, -\dvert)$) {};\node[left=1.123mm of X9,nodelbl] {$\Gamma$};
      \node[fold] (Z13) at ($(Y12) + (-.5, -\dvert)$) {};\node[right=0.4789mm of Z13,nodelbl] {$\Phi$};
      \node[showdown] (Z14) at ($(Y12) + (.5, -\dvert)$) {};\node[right=0.4789mm of Z14,nodelbl] {$\Theta$};
      \node[fold] (Z10) at ($(X9) + (-.5, -\dvert)$) {};\node[left=0.4789mm of Z10,nodelbl] {$\Psi$};
      \node[showdown] (Z11) at ($(X9) + (.5, -\dvert)$) {};\node[right=0.4789mm of Z11,nodelbl] {$\Omega$};
      \node[showdown] (Z23) at ($(X21) + (-1.2, -\dvert)$) {};\node[left=0.4789mm of Z23,nodelbl] {\textsc{t}};
      \node[pl2] (Y24) at ($(X21) + (0, -\dvert)$) {}; \node[left=1.123mm of Y24,nodelbl] {\textsc{u}};
      \node[fold] (Z22) at ($(X21) + (1.2, -\dvert)$) {};\node[left=0.4789mm of Z22,nodelbl] {\textsc{v}};
      \node[fold] (Z28) at ($(X27) + (-.5, -\dvert)$) {};\node[below=.5mm of Z28,nodelbl] {\textsc{w}};
      \node[showdown] (Z29) at ($(X27) + (.5, -\dvert)$) {};  \node[below=.5mm of Z29,nodelbl] {\textsc{x}};
      \node[fold] (Z25) at ($(Y24) + (-.5, -\dvert)$) {};\node[right=0.4789mm of Z25,nodelbl] {$\Lambda$};
      \node[showdown] (Z26) at ($(Y24) + (.5, -\dvert)$) {};\node[right=0.4789mm of Z26,nodelbl] {$\Xi$};

      \draw [edg,->] (X0) -- (Y1) node[pos=.7,acsty] {\fontsize{\acsz}{\acsz}\selectfont check} node[pos=\sqpos,sqsty]{\small\seq{1}};
      \draw [edg,->] (X0) -- (Y30) node[pos=.6,acsty] {\fontsize{\acsz}{\acsz}\selectfont all~in} node[pos=\sqpos,sqsty]{\small\seq{2}};
      \draw [edg,->] (X0) -- (Y18) node[pos=.7,acsty] {\fontsize{\acsz}{\acsz}\selectfont bet} node[pos=\sqpos,sqsty]{\small\seq{3}};
      \draw [edg,dashed,->] (Y1) -- (X3) node[pos=\acpos,acsty] {\fontsize{\acsz}{\acsz}\selectfont bet} node[pos=\sqpos,sqsty]{\small\seq{1}};
      \draw [edg,dashed,->] (Y1) -- (Z2) node[pos=\acpos,acsty] {\fontsize{\acsz}{\acsz}\selectfont check} node[pos=\sqpos,sqsty]{\small\seq{2}};
      \draw [edg,dashed,->] (Y1) -- (X15) node[pos=\acpos,acsty,xshift=1mm] {\fontsize{\acsz}{\acsz}\selectfont all~in} node[pos=\sqpos,sqsty]{\small\seq{3}};
      \draw [edg,dashed,->] (Y30) -- (Z31)  node[pos=\acpos,acsty] {\fontsize{\acsz}{\acsz}\selectfont fold} node[pos=\sqpos,sqsty]{\small\seq{9}};
      \draw [edg,dashed,->] (Y30) -- (Z32) node[pos=\acpos,acsty] {\fontsize{\acsz}{\acsz}\selectfont call} node[pos=\sqpos,sqsty]{\small\seq{10}};
      \draw [edg,dashed,->] (Y18) -- (Z20) node[pos=\acpos,acsty] {\fontsize{\acsz}{\acsz}\selectfont call} node[pos=\sqpos,sqsty]{\small\seq{11}};
      \draw [edg,dashed,->] (Y18) -- (X21) node[pos=\acpos,acsty] {\fontsize{\acsz}{\acsz}\selectfont raise} node[pos=\sqpos,sqsty]{\small\seq{12}};
      \draw [edg,dashed,->] (Y18) -- (Z19) node[pos=\acpos,acsty] {\fontsize{\acsz}{\acsz}\selectfont fold} node[pos=\sqpos,sqsty]{\small\seq{13}};
      \draw [edg,dashed,->] (Y18) -- (X27) node[pos=\acpos,acsty] {\fontsize{\acsz}{\acsz}\selectfont all~in} node[pos=\sqpos,sqsty]{\small\seq{14}};
      \draw [edg,->] (X3) -- (Z5) node[pos=\acpos,acsty] {\fontsize{\acsz}{\acsz}\selectfont call} node[pos=\sqpos,sqsty]{\small\seq{4}};
      \draw [edg,->] (X3) -- (Y6) node[pos=\acpos,acsty] {\fontsize{\acsz}{\acsz}\selectfont raise} node[pos=\sqpos,sqsty]{\small\seq{5}};
      \draw [edg,->] (X3) -- (Z4) node[pos=\acpos,acsty] {\fontsize{\acsz}{\acsz}\selectfont fold} node[pos=\sqpos,sqsty]{\small\seq{6}};
      \draw [edg,->] (X3) -- (Y12) node[pos=\acpos,acsty] {\fontsize{\acsz}{\acsz}\selectfont all~in} node[pos=\sqpos,sqsty]{\small\seq{7}};
      \draw [edg,->] (X15) -- (Z16) node[pos=\acpos,acsty] {\fontsize{\acsz}{\acsz}\selectfont fold} node[pos=\sqpos,sqsty,xshift=-.3mm]{\small\seq{10}};
      \draw [edg,->] (X15) -- (Z17) node[pos=\acpos,acsty] {\fontsize{\acsz}{\acsz}\selectfont call} node[pos=\sqpos,sqsty,xshift=.3mm]{\small\seq{11}};
      \draw [edg,dashed,->] (Y6) -- (Z7) node[pos=\acpos,acsty] {\fontsize{\acsz}{\acsz}\selectfont fold} node[pos=\sqpos,sqsty]{\small\seq{4}};
      \draw [edg,dashed,->] (Y6) -- (Z8) node[pos=\acpos,acsty] {\fontsize{\acsz}{\acsz}\selectfont call} node[pos=\sqpos,sqsty]{\small\seq{5}};
      \draw [edg,dashed,->] (Y6) -- (X9)node[pos=\acpos,acsty] {\fontsize{\acsz}{\acsz}\selectfont all~in} node[pos=\sqpos,sqsty]{\small\seq{6}};
      \draw [edg,dashed,->] (Y12) -- (Z13) node[pos=\acpos,acsty] {\fontsize{\acsz}{\acsz}\selectfont fold} node[pos=\sqpos,sqsty]{\small\seq{7}};
      \draw [edg,dashed,->] (Y12) -- (Z14) node[pos=\acpos,acsty] {\fontsize{\acsz}{\acsz}\selectfont call} node[pos=\sqpos,sqsty]{\small\seq{8}};
      \draw [edg,->] (X9) -- (Z10) node[pos=\acpos,acsty] {\fontsize{\acsz}{\acsz}\selectfont fold} node[pos=\sqpos,sqsty]{\small\seq{8}};
      \draw [edg,->] (X9) -- (Z11) node[pos=\acpos,acsty] {\fontsize{\acsz}{\acsz}\selectfont call} node[pos=\sqpos,sqsty]{\small\seq{9}};
      \draw [edg,->] (X21) -- (Z23) node[pos=\acpos,acsty] {\fontsize{\acsz}{\acsz}\selectfont call} node[pos=\sqpos,sqsty]{\small\seq{12}};
      \draw [edg,->] (X21) -- (Y24) node[pos=\acpos,acsty] {\fontsize{\acsz}{\acsz}\selectfont all~in} node[pos=\sqpos,sqsty]{\small\seq{13}};
      \draw [edg,->] (X21) -- (Z22) node[pos=\acpos,acsty] {\fontsize{\acsz}{\acsz}\selectfont fold} node[pos=\sqpos,sqsty]{\small\seq{14}};
      \draw [edg,->] (X27) -- (Z28)node[pos=\acpos,acsty] {\fontsize{\acsz}{\acsz}\selectfont fold} node[pos=\sqpos,sqsty,xshift=-.3mm]{\small\seq{15}};
      \draw [edg,->] (X27) -- (Z29)  node[pos=\acpos,acsty] {\fontsize{\acsz}{\acsz}\selectfont call} node[pos=\sqpos,sqsty,xshift=.3mm]{\small\seq{16}};
      \draw [edg,dashed,->] (Y24) -- (Z25) node[pos=\acpos,acsty] {\fontsize{\acsz}{\acsz}\selectfont fold} node[pos=\sqpos,sqsty,xshift=-.3mm]{\small\seq{15}};
      \draw [edg,dashed,->] (Y24) -- (Z26) node[pos=\acpos,acsty] {\fontsize{\acsz}{\acsz}\selectfont call} node[pos=\sqpos,sqsty,xshift=.3mm]{\small\seq{16}};
    \end{tikzpicture}}\vspace{-2mm}
  \caption{Skeleton of a river endgame. The initial pot contributions for the endgame are $\$1875$ for both players. Each player has a stack size worth $\$18125$ they can play. Every `bet' and `raise' action corresponds to first matching the other player's contribution to the pot, and then increasing the player's contribution to the pot by an amount equal to $\nicefrac{3}{4}$ of the cumulative amount in the pot.
    Black nodes belong to the small blind player, white nodes to the big blind player. The symbol \protect\tikz{\protect\node[showdown] at (0,0) {};} denotes a showdown, while \protect\tikz{\protect\node[fold] at (0,0) {};} denotes that one player folded. 
  }\vspace{-2mm}
  \label{fig:river skeleton appendix}
\end{figure*}

\subsection{Computation of pot contributions and stack amounts}
As an illustration, consider state
\textsc{d} in the game of \cref{fig:river skeleton}, which is the state reached
from the root \textsc{a} when the small blind player first matches the
contribution of the big blind player, and then further raises their
contribution by an amount equal to $\nicefrac{3}{4}$ of the pot. Since the pot
contributions of the two players at state $\textsc{a}$ are equal (to $\$1875$),
the small blind player has already matched the contribution, and moves
$\nicefrac{3}{4}(\$1875\times 2) = \$2812.5$ from their stack into the pot. So,
at \textsc{d} the stacks of the players are $\$15312.5\ (=\$18125-\$2812.5)$
for the small blind player and $\$18125$ for the big blind player, and the
contributions to pot are $\$4687.5\ (=\$1875 + \$2812.5)$ for the small blind
player and $\$1875$ for the big blind player.
The complete list of pot
contributions at each terminal state of the skeleton of \cref{fig:river skeleton} is shown on the right of the skeleton.

    \section{Sparsification techniques}
We will find the following property of the Kronecker product useful.

\begin{property}
    The Kronecker product operation is bilinear and associative but not
    commutative.
\end{property}
\begin{property}\label{pro:kro transpose}
    Transposition distributes over Kronecker
    products: $(\vec{P}\kro\vec{Q})^\top = \vec{P}^\top \kro \vec{Q}^\top.$
\end{property}
\begin{property}\label{pro:kro inverse}
    Inversion distributes over Kronecker
    products: $(\vec{P}\kro\vec{Q})^{-1} = \vec{P}^{-1} \kro \vec{Q}^{-1}.$
\end{property}
\begin{property}[Mixed-product rule]\label{pro:mixed product}
    Let $\vec{P}\in\bbR^{m\times n},\vec{Q}\in\bbR^{r\times s},\vec{C}\in\bbR^{n\times \ell},\vec{D}\in\bbR^{s\times q}$ be arbitrary matrices. Then, $(\vec{P}\vec{C})\kro(\vec{Q}\vec{D}) = (\vec{P}\kro\vec{Q})(\vec{C}\kro\vec{D}).$
\end{property}

\subsection{More details on Technique A}
The first technique sparsifies the term $(\vec{\Lambda}_1\vec{W}\vec{\Lambda}_2)\kro\vec{S}$ by
recursively sparsifying the `win-lose' matrix $\vec{W}$.
Specifically, it first computes a
sparsification $\vec{W} = \hat{\vec{W}} + \vec{U}_W\vec{V}_W^\top$ (in our experiments,
we do so by using the general heuristic described in \citep{Zhang20:Sparsified}). With that,
\begin{align}
    (\vec{\Lambda}_1\vec{W}\vec{\Lambda}_2)\kro \vec{S} & = \Big(\vec{\Lambda}_1(\hat{\vec{W}} + \vec{U}_W \vec{V}_W^\top)\vec{\Lambda}_2\Big)\kro \vec{S}\nonumber                                                                         \\
                                                        & = (\vec{\Lambda}_1\hat{\vec{W}}\vec{\Lambda}_2)\kro \vec{S} + \Big((\vec{\Lambda}_1\vec{U}_W) (\vec{V}_W^\top\vec{\Lambda}_2)\Big)\kro(\vec{I}\vec{S})\nonumber                   \\
                                                        & = (\vec{\Lambda}_1\hat{\vec{W}}\vec{\Lambda}_2) \kro \vec{S} + \Big((\vec{\Lambda}_1\vec{U}_W) \kro \vec{I}\Big) \Big((\vec{\Lambda}_2\vec{V}_W) \kro \vec{S}^\top\Big)^{\!\top},
\end{align}
where the second equality follows from the bilinearity of Kronecker product, and
the last equality follows from the mixed-product rule and \cref{pro:kro transpose}.

\subsection{More details on Technique B}
The second technique leverages the fact that the hands of each player can be ranked
by their strength. When that is done (ignore for now incompatible hands) each row of the win-lose matrix $\vec{W}$ begins with zero or more
columns equal to $-1$, followed by zero or more columns with value $0$, followed by zero or more columns with
value $1$. As the hand of the row player becomes stronger, the number of $-1$'s on the row decreases, and the number of $1$'s increases. Hence, the matrix obtained by subtracting from each row of $\vec{W}$ the previous line must be very sparse. We can
compactly represent the operation of subtracting from each row of $\vec{W}$ the preceding row via the matrix operation $\vec{Y} \defeq \vec{D}\vec{W}$,
where the lower bidiagonal matrix
\[
    \vec{D} = \begin{bmatrix}
        1  &                     \\
        -1 & 1                   \\
           & \ddots & \ddots     \\
           &        & -1     & 1
    \end{bmatrix}
\]
has value $1$ on the main diagonal, and value $-1$ in the diagonal below the main diagonal. Then,
\begin{align}
    (\vec{\Lambda}_1\vec{W}\vec{\Lambda}_2)\kro \vec{S} & = \Big(\vec{\Lambda}_1(\vec{D}^{-1} \vec{Y})\vec{\Lambda}_2\Big)\kro (\vec{I}\vec{S})\nonumber                                                      \\
                                                        & = (\vec{\Lambda}_1\kro \vec{I})\Big((\vec{D}^{-1}(\vec{Y}\vec{\Lambda}_2)) \kro \vec{S}\Big)\nonumber                                               \\
                                                        & = (\vec{\Lambda}_1\kro \vec{I})\Big((\vec{D}^{-1}(\vec{Y}\vec{\Lambda}_2)) \kro (\vec{I}\vec{S})\Big)\nonumber                                      \\
                                                        & = (\vec{\Lambda}_1\kro \vec{I})(\vec{D}^{-1}\kro \vec{I})(\vec{Y}\vec{\Lambda}_2) \kro \vec{S})\nonumber                                            \\
                                                        & = (\vec{\Lambda}_1\kro\vec{I})(\vec{D}\kro\vec{I})^{-1}\Big((\vec{\Lambda}_2\vec{Y}^\top) \kro \vec{S}^\top\Big)^{\!\top}\!\!,\label{eq:technB app}
\end{align}
where the used \cref{pro:kro inverse,pro:kro transpose} in the last step.

\section{Libratus' betting abstraction}
The betting abstraction we use in the experiments matches the abstraction used by Libratus. Specifically, the player corresponding to the AI has the following bet amounts (we omit all-ins, which apply to all bullet points):
\begin{itemize}[nolistsep,itemsep=1mm,leftmargin=8mm]
    \item As the first action of the river endgame: bets worth 0.25x, 0.5x, 1x, 2x, 4x, 8x of the pot amount;
    \item When faced with a check: 0.25x, 0.5x, 1x, 2x, 4x, 8x of the pot amount;
    \item When faced with an initial bet (no raises yet): 0.4x, 0.7x, 1.1x, 2x of the pot amount;
    \item When there has been a single raise: 0.4x, 0.7x, 2x of the pot amount;
    \item When faced with subsequent raises: 0.7x of the pot amount.
\end{itemize}

The opponent uses the following bet amounts (we again omit all-ins, which apply to all bullet points):
\begin{itemize}[nolistsep,itemsep=1mm,leftmargin=8mm]
    \item As the first action of the river endgame: 0.35x, 0.65x, 1x of the pot amount;
    \item When faced with a check: 0.5x, 0.75x, 1x of the pot amount;
    \item When faced with an initial bet (no raises yet): 0.7x, 1.1x of the pot amount;
    \item When faced with subsequent raises: 0.7x of the pot amount.
\end{itemize}

There is no cap on the number of raises per round.



\fi

\end{document}